\documentclass{iopart}

\usepackage{mathptmx}
\usepackage{graphicx}
\usepackage{iopams}
\usepackage{setstack}
\usepackage{amsopn}

\newcommand{\ix}[1]{\ensuremath{\mathrm{#1}}}
\DeclareMathOperator{\Real}{Re}
\DeclareMathOperator{\Imag}{Im}

\begin{document}

%====================================================================

%====================================================================

\title{Properties of multi-particle Green and vertex functions within
  Keldysh formalism}

\author{Severin G Jakobs, Mikhail Pletyukhov and Herbert Schoeller}

\address{Institut f\"ur Theoretische Physik~A, RWTH Aachen
  University, D-52056 Aachen, Germany}

\address{JARA -- Fundamentals of Future Information Technology, Aachen
  and J\"ulich, Germany}

\ead{sjakobs@physik.rwth-aachen.de}

\begin{abstract}
  The increasing interest in nonequilibrium effects in condensed
  matter theory motivates the adaption of diverse equilibrium
  techniques to Keldysh formalism. For methods based on multi-particle
  Green or vertex functions this involves a detailed knowledge of the
  real-time properties of those functions. In this review, we derive
  general properties of fermionic and bosonic multi-particle Green and
  vertex functions for a stationary state described within Keldysh
  formalism. Special emphasis is put on the analytic properties
  associated with causality and on a detailed discussion of the
  Kubo-Martin-Schwinger conditions which characterize thermal
  equilibrium. Finally we describe how diagrammatic approximations and
  approximations within the functional renormalization group approach
  respect these properties.
\end{abstract}


\maketitle

%====================================================================

%====================================================================

\section{Introduction}
\label{sec:introduction}

Multi-particle Green and vertex functions are an important tool for
the theoretical analysis of interacting quantum systems in condensed
matter physics. $n$-particle Green functions contain the relevant
information to compute $n$-particle properties of the
system~\cite{negele}. Moreover, determining multi-particle functions
may be an auxiliary step on the way to find the single-particle Green
function (or an approximation to it) and single-particle properties.
For example, the functional renormalization group (fRG)
approach~\cite{salmhofer1} leads to a system of flow equations which
couples all $n$-particle functions with $n \ge 1$ in an infinite
hierarchy. Depending on the particular fRG scheme, these $n$-particle
functions may be Green or irreducible vertex functions.  Another
example is given by diagrammatic resummation techniques based on the
Bethe-Salpeter equations, such as fluctuation exchange~\cite{bickers1}
or parquet-equations~\cite{dominicis, bickers2}. For these approaches
the preliminary task is to determine the irreducible two-particle
vertex function, from which in turn the single-particle self-energy
can be deduced.

The methods mentioned above are typically used within zero-temperature
or equilibrium formalism. Calculations in equilibrium formalism are
based on imaginary frequencies and require an analytic continuation to
real frequencies in order to determine many physical observables of
interest. At finite temperature, such analytic continuation is
numerically quite nontrivial even for a single frequency (see
e.g.~\cite{beach} and references therein). Therefore a formulation of
the problem within the real-time (real-frequency) Keldysh
formalism~\cite{rammer, haug, lifshitz} may be advantageous. Beyond
that, Keldysh formalism allows for the description of nonequilibrium
situations such as stationary state transport at finite bias voltage.

A central ingredient of Keldysh formalism is a time contour consisting
of two branches, one of them describing evolution forward and the
other backward in time. As a consequence of the existence of these two
branches, the $n$-particle Green function is a tensor in
$2$-dimensional space of rank $2n$, having $2^{2n}$ components, each
being a function of $2n$ state indices and $2n$ frequencies [or
$(2n-1)$ frequencies in the case of time translational invariance].
The resulting additional complexity of the formalism is however
reduced due to the fact that the components are not independent from
each other but obey certain relations connecting them. In this review,
we give a comprehensive and self-contained presentation of fundamental
relations for the multi-particle Green and vertex functions within
Keldysh formalism: the behaviour under exchange of particles and under
complex conjugation, properties connected to causality and the
Kubo-Martin-Schwinger (KMS) conditions characterizing thermal
equilibrium. The special single-particle versions of these relations
are commonly known, e.g. the fact that retarded and advanced
single-particle Green function are adjoint to each other, and the
fluctuation dissipation theorem in thermal equilibrium. However, also
the general multi-particle versions are required to adapt the
above-mentioned methods to Keldysh formalism, for instance for real
time implementations of the fRG, which have been initiated
in~\cite{gezzi, jakobs1}.

Some of the properties in question have been analysed in~\cite{chou,
  wang} for multi-point Green functions based on real bosonic
fields. The adaption of their results to condensed matter problems is
far from being obvious due to two major complications. Firstly,
condensed matter problems usually involve complex bosonic or Grassmann
fermionic operators instead of real bosonic ones. This makes necessary
a separate book-keeping of incoming and of outgoing indices
(corresponding to annihilation and creation operators) instead of just
one type of indices as for real bosons. Additionally, due to the
commutation relations of complex bosonic or Grassmann fermionic
operators, the initial order of a time-ordered operator product is
relevant for the result, as opposed to real bosons. Furthermore, the
time reversal properties of the field operators which are important
for the analysis of the KMS conditions differ from those of real
bosonic field operators. The second source of complication is due to
the fact that condensed matter problems usually involve external
potentials which lift diverse symmetries of the
system. Reference~\cite{wang} implicitly assumes translational
invariance in space and time and invariance under the product of time
reversal and parity. These prerequisites, which considerably simplify
the investigation of the KMS conditions, are not given in typical
condensed matter problems.  Despite these fundamental differences some
of the concepts developed in references~\cite{chou, wang} can be
fruitfully adapted to our analysis.

In section~\ref{sec:system}, we start by a short description of the
physical situation under consideration. The set-up of an interacting
quantum system coupled to a number of noninteracting reservoirs
represents a typical stationary state configuration described in
Keldysh formalism. 

We continue by specifying precisely the definitions of the Green and
vertex functions in question in section~\ref{sec:def}. Here it is
necessary to take care of the details (like prefactor conventions)
since the exact form of the relations to be discussed depends on them.

In sections~\ref{subsec:permutation} and~\ref{subsec:conjugation} we
derive useful relations connected to the permutation of particles and
to complex conjugation.  Section~\ref{subsec:causality} focuses on a
theorem of causality~\cite{chou} which we cast in concise form. We
show that this theorem has direct consequences for the analytic
properties of the Green and vertex functions in Fourier space and
identify components which are analytic in one half plane of each of
their $(2n-1)$ independent frequency arguments. Remarkably, our
derivation does not use the Lehmann representation. This fact is
highly advantageous due to the unwieldy form the latter acquires in
the multiparticle case.

Section~\ref{sec:equilibrium} is devoted to the special situation of
thermal equilibrium. The fact that the density matrix then corresponds
to a time evolution in imaginary time is at the root of the KMS
conditions~\cite{kubo, martin}.  In section~\ref{subsec:KMS}, we
describe how to concisely formulate the multi-particle KMS conditions
exploiting the so-called tilde Green function which is defined on the
analogy of~\cite{wang} by means of a non-standard ordering of
operators on the Keldysh contour.  However, in order to extract useful
information from the KMS conditions, the tilde Green function has to
be eliminated in favour of the standard Green function. In
sections~\ref{subsec:time_reversal} and~\ref{subsec:gen_fluc_diss} we
find that time reversal can be used for this purpose, an observation
similar to the one of~\cite{chou} for multi-point Green functions
based on real bosons.  In section~\ref{subsec:time_rev_inv_GF} we
characterize an important class of systems (including e.g. the
Anderson impurity model~\cite{anderson}) with specific time reversal
properties.  For this class, we formulate a multi-particle FDT in
section~\ref{subsec:gen_fluc_diss_spec}.

Finally we study under which conditions approximations to the Green or
vertex functions conserve the before-mentioned properties
(section~\ref{sec:approx}). These conditions turn out to be quite
handy for the large class of diagrammatic approximations. For
approximations within the fRG framework, the choice of the flow
parameter is found to be crucial for the conservation of causal
features and of the KMS conditions.

%====================================================================

%====================================================================

\section{The system under consideration}
\label{sec:system}

The physical setup underlying our considerations is a finite
interacting quantum system of a single kind of either bosonic or
fermionic particles which is coupled to $M$ noninteracting reservoirs,
$M \ge 1$. Let $\{q\} = \{s\} \cup \{r_1\} \cup \ldots \cup \{r_M\}$
be a basis of single-particle states which are either localized in the
system (states $s$) or in one of the reservoirs (states $r_k$, $k = 1,
\dots, M$). We use the index $q$ to refer to any of those states. The
total Hamiltonian under consideration is given by
\begin{equation}
  \label{eq:full_H}
  H = H_\ix{sys} + \sum_{k=1}^M H^{(k)}_\ix{res} + \sum_{k=1}^M
  H^{(k)}_\ix{coup},
\end{equation}
with
\begin{eqnarray}
  \label{eq:H_sys}
  H_\ix{sys}
  =
  \sum_{s, s'} h_{s'|s}
  a^\dagger_{s'} a_{s}
  +
  \frac{1}{4} \sum_{\substack{s_1, s'_1\cr s_2, s'_2}}
  \overline{v}_{s'_1 s'_2| s_1 s_2}
  a^\dagger_{s'_1}a^\dagger_{s'_2} a_{s_2} a_{s_1},
  \\
  H^{(k)}_\ix{res}
  =
  \sum_{r_k, r'_k} h_{r'_k|r_k}
  a^\dagger_{r'_k} a_{r_k},
  \\
  H^{(k)}_\ix{coup}
  =
  \sum_{s, r_k} h_{s|r_k}  a^\dagger_{s} a_{r_k}
  +
  \mbox{H.c.}
\end{eqnarray}
in standard notation of second quantitation. Here
\begin{equation}
  h_{q'|q} = \langle q'|h|q \rangle = h_{q|q'}^\ast
\end{equation}
are the matrix elements of the single-particle Hamiltonian and
\begin{equation}
  \label{eq:vertex_def}
  \overline{v}_{s'_1 s'_2| s_1 s_2} =
  \langle s'_1 s'_2| v | s_1 s_2 \rangle
  + \zeta \langle s'_1 s'_2| v | s_2 s_1 \rangle
\end{equation}
denote the (anti-)symmetrized matrix elements of the two-particle
interaction, with
\begin{equation}
  \zeta = 
  \cases{+ 1 &for bosons,\\ - 1 &for fermions.}
\end{equation}

At some initial time $t_0$ the whole configuration is assumed to be
described by the density matrix
\begin{equation}
  \rho(t_0)
  =
  \rho_\ix{sys}
  \otimes
  \rho_\ix{res}^{(1)}
  \otimes
  \ldots
  \otimes
  \rho_\ix{res}^{(M)},
\end{equation}
where
\begin{equation}
  \label{eq:rho_res}
  \rho_\ix{res}^{(k)} = \frac{\rme^{- \beta_k (H_k - \mu_k N_k)}}{\Tr \rme^{-
      \beta_k (H_k - \mu_k N_k)}},
  \qquad k = 1, \ldots, M,
\end{equation}
is the grand-canonical distribution of reservoir $k$ with temperature
$T_k = 1/\beta_k$ and chemical potential $\mu_k$. Keldysh formalism
allows us to compute properties described by the density matrix
$\rho(t)$, $t > t_0$, which evolves unitarily from $\rho(t_0)$ under
the full Hamiltonian $H$ given in~\eref{eq:full_H} \cite{rammer, haug,
  lifshitz}.

In the limit of an infinite reservoir size $L$ and an infinite
transient time $t-t_0$ such that $L/u \gg -t_0 \rightarrow \infty$
(with $u$ being the particle velocity in the reservoir), expectation
values under $\rho(t)$ become stationary and do not depend on the
initial density matrix $\rho_\ix{sys}$ of the finite interacting
region. We will restrict our considerations to this stationary
situation.

A particular stationary situation is a global equilibrium which occurs
if all $T_k$ and $\mu_k$ are equal. We specialism to the equilibrium
case in section~\ref{sec:equilibrium}. The results found there are
also applicable to interacting bulk systems (that is large systems
with the Hamiltonian~\eref{eq:H_sys}) in thermal equilibrium.

%====================================================================

%====================================================================

\section{Definition of the Green and vertex functions}
\label{sec:def}

\begin{figure}
  \centering
  \includegraphics{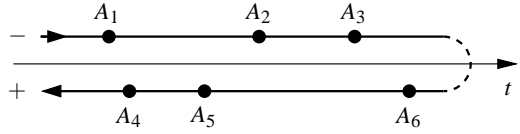}
  \caption{The time contour consisting of the $-$- and the $+$-branch
    with some example operators $A_1, \ldots, A_6$.  The time ordering
    operator $T_\ix{c}$ will establish the order $A_4 A_5 A_6 A_3 A_2
    A_1$. The operator $\widetilde T_\ix{c}$ used in the
    definition~\eref{eq:tilde_GF} of the tilde Green function will
    establish the order $A_6 A_5 A_4 A_1 A_2 A_3$.}
    \label{fig:contour}
\end{figure}

The time- or frequency-dependent $n$-particle Green function is
defined by
\begin{eqnarray}
  \label{eq:GF}
  G^{j|j'}_{q|q'}(t|t') =
  (-\rmi)^n \Big \langle T_\ix{c} \, a_{q_1}^{(j_1)}(t_1) \ldots
  a_{q_n}^{(j_n)}(t_n) a_{q'_n}^{(j'_n)\,\dagger}(t'_n) \ldots
  a_{q'_1}^{(j'_1)\, \dagger}(t'_1) \Big \rangle,
  \\
  \label{eq:GF_freq}
  G^{j|j'}_{q|q'}(\omega|\omega') = \int \! \rmd t_1 \ldots \rmd t_n' \,
  \rme^{\rmi(\omega \cdot t - \omega' \cdot t')} G^{j|j'}_{q|q'}(t|t'),
\end{eqnarray}
where $q = (q_1, \ldots, q_n)$, $t = (t_1, \ldots, t_n)$, $\omega =
(\omega_1, \ldots, \omega_n)$ and $j = (j_1,\ldots, j_n)$ are
multi-indices denoting state, time, frequency and branch of the time
contour, respectively. The annihilation and creation operators
$a_{q_k}$ and $a^\dagger_{q_k}$ of a particle in the state $q_k$ obey
standard commutation relations
\begin{eqnarray}
  [a_{q_1}, a^\dagger_{q_2}]_{-\zeta}
  =
  a_{q_1} a^\dagger_{q_2} - \zeta a^\dagger_{q_2} a_{q_1}
  =
  \delta_{q_1,q_2},
  \label{eq:comm_rel}
  \\
  \label{eq:comm_relA}
  [a_{q_1}, a_{q_2}]_{-\zeta}
  =
  [a^\dagger_{q_1}, a^\dagger_{q_2}]_{-\zeta}
  = 
  0,
\end{eqnarray}
and $a^{(j_k)}_{q_k}(t_k)$ denotes an annihilation operator at time
$t_k$ in the Heisenberg picture with reference time $t_0$. The
operator is situated on the branch $j_k = \mp$ of the time
contour~\cite{rammer}.  The contour ordering operator $T_\ix{c}$
in~\eref{eq:GF} rearranges the sequence of operators in the following
way: all operators with $+$-index are placed left of all operators
with $-$-index. The block of operators with $+$-index is anti-time
ordered internally (which means time arguments of the $+$-operators
increase from left to right), whereas the block of operators with
$-$-index is time ordered (which means time arguments of the
$-$-operators decrease from left to right). By definition of
$T_\ix{c}$ the expression ordered in this way has to be multiplied by
a factor $\zeta$, if it represents an odd permutation of the initial
order given in~\eref{eq:GF}.  The branches of the contour are depicted
in figure~\ref{fig:contour} where also an example for the order
established by $T_\ix{c}$ is given. The expectation value
in~\eref{eq:GF} is defined by
\begin{equation}
  \label{eq:expectation_value}
  \langle A(t) \rangle
  =
  \Tr \rho(t_0) A(t),
\end{equation}
with $\rho(t_0)$ being the density operator at time $t_0 \rightarrow -
\infty$. In~\eref{eq:GF_freq} we used the shorthand notation
$\omega \cdot t = \omega_1 t_1 + \ldots + \omega_n t_n$.

Due to the time translational invariance of the stationary state, the
$n$-particle Green function can be expressed as a function of only
$(2n-1)$ independent time or frequency arguments. For example $t_1$
and $\omega_1$ can be eliminated by
\begin{eqnarray}
  G(t_1, \ldots, t_n|t'_1, \ldots, t'_n)
  =
  G(0, t_2-t_1, \ldots, t_n-t_1|t'_1-t_1, \ldots, t'_n-t_1),
  \\
  \nonumber
  G(\omega_1, \ldots, \omega_n|\omega'_1, \ldots, \omega'_n) =
  2 \pi \delta(\omega_1 + \ldots + \omega_n - \omega'_1 - \ldots -
  \omega'_n)
  \\
  \label{eq:freq_cons}
    \hspace{14em}
  \times
  G(t_1=0, \omega_2, \ldots, \omega_n|\omega'_1, \ldots,
  \omega'_n).
\end{eqnarray}
However, many of the properties discussed in the following are
formulated most easily when all time and frequency arguments are
treated on equal footing. Therefore we will use the reduced number of
frequency arguments only occasionally, e.g. for the study of analytic
properties in section~\ref{subsec:causality}.

Apart from Green functions we will also discuss the properties of
one-particle irreducible vertex functions. These vertex functions can
be derived from a generating functional called effective action which
is the Legendre transform of the generating functional for the
connected Green functions. Details on this path-integral approach to
the irreducible vertex functions can be found for Matsubara formalism
e.g. in~\cite{negele} and for Keldysh formalism in~\cite{gezzi}. Since
we do not need any recourse to the generating functionals in the
following, we choose an equivalent way to introduce the vertex
functions which is based on a diagrammatic expansion.

To set up a standard Hugenholtz diagrammatic technique~\cite{negele}
for a problem with instantaneous two-particle interaction $v$, we
define the (anti-)symmetrized interaction vertex
\begin{equation}
  \fl
  \label{eq:bare_vertex_time}
  \bar v^{j_1' j_2'|j_1 j_2}_{q_1' q_2'|q_1 q_2}(t_1', t_2'|t_1, t_2)
  =
  \delta(t_1 = t_2 = t_1' = t_2')
  (-j_1) \delta_{j_1 = j_2 = j_1' = j_2'}
  \overline{v}_{q'_1 q'_2| q_1 q_2}
\end{equation}
or, in Fourier space,
\begin{eqnarray}
  \fl
  \label{eq:bare_vertex}
  \bar v^{j_1' j_2'|j_1 j_2}_{q_1' q_2'|q_1 q_2}(\omega_1',
  \omega_2'|\omega_1, \omega_2) = 2 \pi \delta(\omega_1 + \omega_2 -
  \omega_1' - \omega_2')
  (-j_1) \delta_{j_1 = j_2 = j_1' = j_2'} 
  \overline{v}_{q'_1 q'_2| q_1 q_2},
\end{eqnarray}
with $\overline{v}_{q'_1 q'_2| q_1 q_2}$ given
by~\eref{eq:vertex_def}.  Since a single Hugenholtz vertex
incorporates both the direct and exchange Feynman vertices, this
diagrammatic language is especially concise in formulation. According
to Wick's theorem, the $n$-particle Green function can be expanded
into a sum of diagrams which consist of lines representing the
noninteracting one-particle Green function $G^{(0)}$ and of
interaction vertices~\cite{danielewicz}.

\begin{figure}
  \centering
  \includegraphics{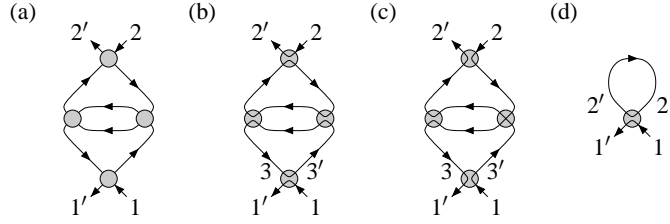}
  \caption{Examples for the application of the diagram rules. The
    lines are free one-particle Green functions, the dots are
    interaction vertices. The indices represent contour index, state
    and frequency, e.g. $1 = (j_1, q_1, \omega_1)$.  Diagram (a)
    contributes to the two-particle vertex function. It has one pair
    of equivalent lines, $n_\ix{eq}=1$, and no symmetries, $S=1$. (b)
    and (c) show two possible, equivalent ways to assign the lines to
    the indices of the vertices. In (b) the vertex at the bottom
    e.g. is $\overline v_{1'3'|13}$, and $n_\ix{loop} = 2$, $\zeta^P =
    1$. In (c) the vertex at the bottom is $\overline v_{1'3'|31} =
    \zeta \overline v_{1'3'|13}$, and $n_\ix{loop} = 0$, $\zeta^P =
    \zeta$. Compared to (b), three of the four vertices acquire a
    factor $\zeta$; taking into account the prefactor
    $\zeta^{n_\ix{loop}} \zeta^P$ the possibilities (b) and (c) hence
    lead to the same value of the diagram. The Hartree-Fock diagram
    (d) ($n_\ix{eq} = 0$, $S = 1$) contributes to the one-particle
    vertex, that is the self-energy. The figure shows the evaluation
    with vertex $\overline v_{1'2'|12}$ and $n_\ix{loop} = 1$.}
    \label{fig:diagr_rule}
\end{figure}

For the discussion of the vertex function we will restrict ourselves
to the frequency representation. The (anti-)symmetrized irreducible
$n$-particle vertex function
\begin{equation}
  \label{eq:vertex}
  \gamma_{q'|q}^{j'|j}(\omega'|\omega)
  =
  \gamma_{q_1' \ldots q_n'|q_1 \ldots q_n}^{j_1' \ldots j_n'|j_1 \ldots
    j_n}(\omega_1', \ldots, \omega_n'|\omega_1, \ldots, \omega_n)
\end{equation}
can be defined diagrammatically as the sum of all one-particle
irreducible diagrams with $n$ amputated incoming lines (having state,
frequency and contour indices $q, \omega, j$) and $n$ amputated
outgoing lines (having indices $q', \omega', j'$). In order to
evaluate a given diagram, one determines first the symmetry factor $S$
which is defined as the number of permutations of vertices mapping the
diagram onto itself. Then one chooses arbitrarily at each vertex the
order in which the attached incoming (outgoing) lines are assigned to
the incoming (outgoing) indices of the vertex. This order should be
kept fixed during the further evaluation. Let $n_\ix{eq}$ count the
pairs of equivalent lines in the diagram, where two lines are called
equivalent if they connect the same vertices and run in the same
direction.  One can define $n_\ix{loop}$ to be the number of internal
loops made up from internal propagators. (Inside a given vertex $\bar
v_{1'2'|12}$ we assume that one line continues from $1$ to $1'$ and
the other continues from $2$ to $2'$.) Finally, the incoming line $k$
being connected to the outgoing line $P(k')$ via internal lines and
vertices determines a permutation $P$ of $\{1', \ldots ,n'\}$. The
value of the diagram is then given by
\begin{equation}
  \label{eq:amp_diagram}
  \frac{\zeta^{n_\ix{loop}} \zeta^{P}}{2^{n_\ix{eq}} S}
  \frac{(2\pi)^{2n}}{\rmi^{n-1}}
  \left[\prod \frac{\rmi}{(2\pi)^4} \bar v \right]
  \prod G^{(0)},
\end{equation}
where one has to integrate over all internal frequencies and sum over
all internal states and contour indices.  Due to the
(anti-)symmetrization of the vertex in~\eref{eq:bare_vertex}, the
overall sign of the diagram is actually independent of the chosen
assignment of lines to indices at each vertex. An example for the
application of the diagram rule is given in
figure~\ref{fig:diagr_rule}. The Hartree-Fock contribution to the
one-particle vertex depicted in figure~\ref{fig:diagr_rule}~(d) for
instance is evaluated as
\begin{equation}
  \fl
  \zeta (2 \pi)^2 \frac{i}{(2 \pi)^4} \sum_{j_2, j_2'} \sum_{q_2, q_2'}
  \int \! \rmd \omega_2 \rmd \omega_2' \, \overline v_{q'_1 q'_2|q_1
    q_2}^{j'_1 j'_2|j_1 j_2}(\omega'_1 \omega'_2|\omega_1 \omega_2)
  {G^{(0)}}_{q_2|q'_2}^{j_2|j'_2}(\omega_2| \omega'_2).
\end{equation}

Given time translational invariance one often works with vertex
functions depending on $(2n-1)$ frequencies only, e.g. $\gamma(t'_1=0,
\omega'_2 \ldots \omega'_n|\omega_1 \ldots \omega_n)$. The full vertex
function~\eref{eq:vertex} is then given by
\begin{eqnarray}
  \fl
  \nonumber
  \gamma(\omega'_1, \ldots, \omega'_n|\omega_1, \ldots, \omega_n) =
  2 \pi \delta(\omega'_1 + \ldots + \omega'_n - \omega_1 - \ldots -
  \omega_n)\\
  \fl
  \hspace{21em}
  \times
  \gamma(t'_1=0, \omega'_2, \ldots, \omega'_n|\omega_1, \ldots,
  \omega_n).
\end{eqnarray}
The diagrams for $\gamma(t'_1=0, \omega'_2 \ldots \omega'_n|\omega_1
\ldots \omega_n)$ typically consist of single-particle Green functions
depending on a single frequency [occurring in the second line
of~\eref{eq:freq_cons}] and of frequency independent vertices [the
part $(-j_1) \delta_{j_1 = j_2 = j_1' = j_2'} \overline{v}_{q'_1 q'_2|
  q_1 q_2}$ of~\eref{eq:bare_vertex}].  When evaluating such diagrams,
the factor $(2\pi)^{2n}$ in~\eref{eq:amp_diagram} has to be replaced
by $(2\pi)^{n-1}$ and $\rmi/(2\pi)^4$ by $\rmi/(2\pi)$.  For the
Hartree-Fock diagram in figure~\ref{fig:diagr_rule}~(d) e.g. this
yields
\begin{equation}
  \zeta \frac{i}{2 \pi} \sum_{j_2, j_2'} \sum_{q_2, q_2'}
  \int \! \rmd \omega_2 \, \overline v_{q'_1 q'_2|q_1
    q_2}^{j'_1 j'_2|j_1 j_2} {G^{(0)}}_{q_2|q'_2}^{j_2|j'_2}(\omega_2).
\end{equation}

Equation~\eref{eq:amp_diagram} defines the prefactor of the vertex
functions in such a way, that they can be used themselves as vertices
in diagrams with the identical prefactor rules being applicable as for
the bare $n$-particle interaction vertices.  The self-energy is equal
to the one-particle vertex, $\Sigma \equiv \gamma_{n=1}$.

In the one-particle case it is common to call the components of the
Green and vertex function chronological, lesser, greater and
antichronological,
\begin{eqnarray} 
  G^\ix{c} = G^{-|-},
  \qquad
  G^< &= G^{-|+},
  \qquad
  G^> &= G^{+|-},
  \qquad
  G^{\widetilde{\ix{c}}} = G^{+|+},
  \\
  \Sigma^\ix{c} = \gamma^{-|-},
  \qquad
  \Sigma^< &= \gamma^{-|+},
  \qquad
  \Sigma^> &= \gamma^{+|-},
  \qquad
  \Sigma^{\widetilde{\ix{c}}} = \gamma^{+|+},
  \label{eq:Sigma}
\end{eqnarray}
respectively.

Apart from the contour basis with indices $j_k = \mp$ we will use the
Keldysh basis~\cite{Keldysh} with indices $\alpha_k = 1,2$ whose
components have a direct physical interpretation in terms of response
and correlations~\cite{hao}. The transformation to the Keldysh basis
is defined by a change of operators,
\begin{equation}
  a^{(1)} 
  = 
  \frac{a^{(-)} - a^{(+)}}{\sqrt{2}},
  \qquad
  a^{(2)} 
  = 
  \frac{a^{(-)} + a^{(+)}}{\sqrt{2}}.
\end{equation}
The Green and vertex functions change according to the tensor
transformation laws,
\begin{eqnarray}
\label{eq:Keldysh_rotation_GF}
  G^{\alpha|\alpha'} 
  &=
  (D^{-1})^{\alpha |j}
  \, G^{j| j'} \,
  D^{j'|\alpha'},
  \\
  \label{eq:Keldysh_rotation_vertex}
  \gamma^{\alpha'|\alpha} 
  &=
  (D^{-1})^{\alpha'|j'}
  \, \gamma^{j'| j} \,
  D^{j|\alpha},
\end{eqnarray}
where an implicit summation over all internal indices is assumed,
and
\begin{eqnarray}
  \label{eq:D}
  D^{j|\alpha} = \prod_{k=1}^n D^{j_k| \alpha_k},
  \\
  \label{eq:D_single}
  D^{-| 1} = D^{\mp| 2} = \frac{1}{\sqrt{2}}, 
  \quad
  D^{+| 1} =  -\frac{1}{\sqrt{2}},
  \\
  \label{eq:D_inv_single}
  (D^{-1})^{1|-} = (D^{-1})^{2|\mp} = \frac{1}{\sqrt{2}}, 
  \quad
  (D^{-1})^{1|+} =  -\frac{1}{\sqrt{2}}.
\end{eqnarray}

In the particular case of one-particle functions, we have
\begin{eqnarray}
  G^{1|1} = 0, 
  \qquad
  G^{1|2} &= G^\ix{Av}, 
  \qquad
  G^{2|1} &= G^\ix{Ret}, 
  \qquad
  G^{2|2} = G^\ix{K},
  \\
  \gamma^{1|1} = \Sigma^\ix{K}, 
  \qquad
  \gamma^{1|2} &= \Sigma^\ix{Ret}, 
  \qquad
  \gamma^{2|1} &= \Sigma^\ix{Av},
  \qquad
  \gamma^{2|2} = 0,
\end{eqnarray}
where the retarded, advanced and Keldysh component of the Green
function and the self-energy are given by
\begin{eqnarray}
  G^\ix{Ret}_{q|q'}(t|t') 
  &=
  - \rmi \Theta(t-t') 
  \Big \langle \big[ a_q(t), a_{q'}^\dagger(t')\big]_{-\zeta} \Big \rangle,
  \\
  G^\ix{Av}_{q|q'}(t|t') 
  &=
   \rmi \Theta(t'-t) 
  \Big \langle \big[ a_q(t), a_{q'}^\dagger(t')\big]_{-\zeta} \Big \rangle,
  \\
  G^\ix{K}_{q|q'}(t|t') 
  &=
  - \rmi \Big \langle \big[ a_q(t), a_{q'}^\dagger(t')\big]_{\zeta} \Big \rangle,
\end{eqnarray}
and
\begin{eqnarray}
  G^\ix{Ret}
  &=
  \label{eq:Dyson_Ret}
  G^{(0) \ix{Ret}}
  +
  G^{(0) \ix{Ret}}
  \circ
  \Sigma^\ix{Ret}
  \circ
  G^\ix{Ret},
  \\
  G^\ix{Av}
  &=
  \label{eq:Dyson_Av}
  G^{(0) \ix{Av}}
  +
  G^{(0) \ix{Av}}
  \circ
  \Sigma^\ix{Av}
  \circ
  G^\ix{Av},
  \\
  G^\ix{K}
  &=
  \label{eq:Dyson_K}
  G^\ix{Ret}
  \circ
  \Sigma^\ix{K}
  \circ
  G^\ix{Av}.
\end{eqnarray}
Here we use ``$\circ$'' as a shorthand for summation over internal
states and integration over internal times,
\begin{equation}
  (A \circ B)_{q|q'}(t|t')
  =
  \sum_r \int \! \rmd s \;
  A_{q|r}(t|s) B_{r|q'}(s|t').
\end{equation}

When discussing the KMS condition in section~\ref{subsec:KMS} we will
need yet another type of Green function, the so-called tilde Green
function which we define on the analogy of~\cite{wang}. It is given by
\begin{eqnarray}
  \label{eq:tilde_GF}
  \widetilde G^{j|j'}_{q|q'}(t|t')
  =
  (-\rmi)^n \Big \langle \widetilde T_\ix{c} \, a_{q_1}^{(j_1)}(t_1) \ldots
  a_{q_n}^{(j_n)}(t_n)
  a_{q'_n}^{(j'_n)\,\dagger}(t'_n) \ldots a_{q'_1}^{(j'_1)\, \dagger}(t'_1)
  \Big \rangle,
  \\
  \label{eq:tilde_GF_freq}
  \widetilde G^{j|j'}_{q|q'}(\omega|\omega')
  =
  \int \! \rmd t_1 \ldots \rmd t_n' \, \rme^{\rmi(\omega \cdot t -
    \omega' \cdot t')} \widetilde G^{j|j'}_{q|q'}(t|t'),
\end{eqnarray}
which differs from~\eref{eq:GF} and~\eref{eq:GF_freq} only
by the other ordering operator $\widetilde T_\ix{c}$. Like $T_\ix{c}$,
$\widetilde T_\ix{c}$ places all operators with $+$-index left of
all operators with $-$-index.  However, under $\widetilde T_\ix{c}$
the block of operators with $+$-index is time ordered internally,
whereas the block of operators with $-$-index is anti-time ordered.
Again, if the final order is an odd permutation of the initial one it
has to be multiplied by a factor $\zeta$.  An example for the order
established by $\widetilde T_\ix{c}$ is given in
figure~\ref{fig:contour}.

For the one-particle case it is straightforward to check the identity
\begin{equation}
  \label{eq:G_G_tilde}
  \widetilde G^{j|j'}_{q|q'}(t|t') = G^{\bar{j'}| \bar j}_{q|q'}(t|t'),
  \quad
  n=1,
\end{equation}
where a bar over a contour index indicates a swap of the branch, $\bar
j = -j$ for $j = \mp$. However, in the general multi-particle case
there is no direct possibility to express the tilde Green function
through an ordinary Green function since the orders established by
$T_\ix{c}$ and $\widetilde T_\ix{c}$ are inherently different; we will
need to make use of time reversal to connect both quantities, see
section~\ref{subsec:time_reversal}.  A direct consequence of having a
special relation for $n=1$ is the particular structure of the
single-particle FDT, see section~\ref{subsec:fluc_diss}.

%======================================================================

%====================================================================

\section{General properties of the Green and vertex functions}
\label{sec:prop}

This section is devoted to properties of the Green and vertex
functions which are valid in a general steady state, whereas
section~\ref{sec:equilibrium} focuses on the particular case of
thermal equilibrium. We will derive the properties of the Green
functions starting out from~\eref{eq:GF} and~\eref{eq:GF_freq}. Once
we know a certain relation for the Green function we can study its
implications for the amputated irreducible diagrams, the values of
which are given by~\eref{eq:amp_diagram}. This will allow us
to infer a corresponding relation for the irreducible vertex
functions.

%--------------------------------------------------------------------

\subsection{Permutation of particles}
\label{subsec:permutation}

Let $P$ be a permutation of $(1, \ldots, n)$ and for any multi-index
$b = (b_1, \ldots, b_n)$ define $Pb := (b_{P(1)}, \ldots, b_{P(n)})$.
The prefactors $\zeta$ due to time ordering in~\eref{eq:GF} and due to
permutations in~\eref{eq:amp_diagram} entail
\begin{eqnarray}
  G^{Pj|j'}_{Pq|q'}(Pt|t')
  =
  G^{j|Pj'}_{q|Pq'}(t|Pt')
  =
  \zeta^P G^{j|j'}_{q|q'}(t|t'),
  \\
  G^{Pj|j'}_{Pq|q'}(P\omega|\omega')
  =
  G^{j|Pj'}_{q|Pq'}(\omega|P\omega')
  =
  \zeta^P G^{j|j'}_{q|q'}(\omega|\omega'),
  \label{eq:G_permute}
  \\
  \gamma^{Pj'|j}_{Pq'|q}(P\omega'|\omega)
  =
  \gamma^{j'|Pj}_{q'|Pq}(\omega'|P\omega)
  =
  \zeta^P \gamma^{j'|j}_{q'|q}(\omega'|\omega).
  \label{eq:gamma_permute}
\end{eqnarray}
The identical equations hold after transforming to the Keldysh basis.
The contour indices $j$ are merely to be replaced by Keldysh indices
$\alpha$.

%--------------------------------------------------------------------

\subsection{Complex conjugation}
\label{subsec:conjugation}

Conjugating~\eref{eq:GF} interchanges creation and
annihilation operators and reverses the order of operators. The
reversed order is identical to the order obtained by $T_\ix{c}$ when
each contour index $j_k$ is swapped to its opposite value $-j_k$.
Therefore, we find
\begin{equation}
  \label{eq:GF(t)_conj}
  G^{j|j'}_{q|q'}(t|t')^\ast
  =
  (-1)^n  G^{\bar {j'}|\bar j}_{q'|q}(t'|t)
\end{equation}
with $\bar j = (\bar j_1, \ldots \bar j_n)$ and $\bar j_k = -j_k$.
Note that the parity of the permutation needed to establish the order
according to $T_\ix{c}$ is identical for $G^{j|j'}_{q|q'}(t|t')$ and
$G^{\bar {j'}|\bar j}_{q'|q}(t'|t)$. Hence no additional sign
prefactors are required in~\eref{eq:GF(t)_conj}. Fourier
transformation leads to
\begin{equation}
  \label{eq:GF_conj_cont}
  G^{j|j'}_{q|q'}(\omega|\omega')^\ast
  =
  (-1)^{n}  G^{\bar {j'}|\bar j}_{q'|q}(\omega'|\omega).
\end{equation}

In order to derive the corresponding identity for the vertex function
we discuss how to conjugate the individual diagrams which the vertex
function is composed of and whose values are given
by~\eref{eq:amp_diagram}. The noninteracting single-particle Green
functions entering the expression~\eref{eq:amp_diagram} can be
conjugated according to~\eref{eq:GF_conj_cont}. The conjugation rule
for the bare two-particle interaction vertices in
expression~\eref{eq:amp_diagram} is inferred from their
definition~\eref{eq:bare_vertex} to be
\begin{equation}
  \label{eq:vert_conj}
  \bar v^{j'|j}_{q'|q}(\omega'|\omega)^\ast
  =
  \bar v^{j'|j}_{q|q'}(\omega'|\omega)
  =
  \bar v^{j|j'}_{q|q'}(\omega|\omega')
  =
  - \bar v^{\bar j|\bar {j'}}_{q|q'}(\omega|\omega'),
\end{equation}
where we make use of $\langle q'_1 q'_2 | v | q_2 q_1 \rangle =
\langle q'_2 q'_1 | v | q_1 q_2 \rangle$ in the first step.
Applying~\eref{eq:GF_conj_cont} and~\eref{eq:vert_conj} to the
conjugate of~\eref{eq:amp_diagram} we find that complex conjugation
maps one-to-one the diagrams contributing to
$\gamma^{j'|j}_{q'|q}(\omega'|\omega)$ onto those of $\gamma^{\bar
  j|\bar {j'}}_{q|q'}(\omega|\omega')$. The precise relationship can
be deduced from the fact that an amputated $n$-particle diagram
comprising $m$ two-particle vertices contains $(2m-n)$ free propagators,
and is found to be
\begin{equation}
  \label{eq:vert_conj_cont}
  \gamma^{j'|j}_{q'|q}(\omega'|\omega)^\ast
  =
  -  \gamma^{\bar j|\bar {j'}}_{q|q'}(\omega|\omega').
\end{equation}

When transforming to the Keldysh basis
via~\eref{eq:Keldysh_rotation_GF}
and~\eref{eq:Keldysh_rotation_vertex} we use that $D$ is real and
fulfils the relation
\begin{equation}
  D^{j|\alpha} = (-1)^{\sum_k \alpha_k} (D^{-1})^{\alpha|\bar j},
\end{equation}
which follows from~\eref{eq:D}~--~\eref{eq:D_inv_single}.
We find
\begin{eqnarray}
  \label{eq:GF_conj}
  G^{\alpha|\alpha'}_{q|q'}(\omega|\omega')^\ast 
  &= 
  (-1)^{n + \sum_k(\alpha_k + \alpha'_k)} 
  \,
  G^{\alpha'|\alpha}_{q'|q}(\omega'|\omega),
  \\
  \label{eq:vertex_conj}
  \gamma^{\alpha'|\alpha}_{q'|q}(\omega'|\omega)^\ast
  &=
  (-1)^{1+\sum_k (\alpha_k + \alpha'_k)}
  \,
  \gamma^{\alpha|\alpha'}_{q|q'}(\omega|\omega').
\end{eqnarray}
For the one-particle functions, equations~\eref{eq:GF_conj}
and~\eref{eq:vertex_conj} yield the well-known relations
\begin{eqnarray}
  \label{eq:GF_conj_1p}
  G^\ix{Ret}_{q|q'}(\omega|\omega') &=
  G^\ix{Av}_{q'|q}(\omega'|\omega)^\ast,
  \\
  G^\ix{K}_{q|q'}(\omega|\omega') &=
  -G^\ix{K}_{q'|q}(\omega'|\omega)^\ast,
\end{eqnarray}
\begin{eqnarray}
  \Sigma^\ix{Ret}_{q|q'}(\omega|\omega') &=
  \Sigma^\ix{Av}_{q'|q}(\omega'|\omega)^\ast,
  \\
  \Sigma^\ix{K}_{q|q'}(\omega|\omega') &=
  -\Sigma^\ix{K}_{q'|q}(\omega'|\omega)^\ast.
\end{eqnarray}

%--------------------------------------------------------------------

\subsection{Causality and analyticity}
\label{subsec:causality}

Consider the Green function $G^{j|j'}(t|t')$ with fixed sets of times
$t = (t_1, \ldots, t_n)$, $t'=(t'_1, \ldots t'_n)$, where $t_1$
happens to be the greatest time argument, $t_1 > t_2, \ldots, t'_n$.
In this case the order of operators established by $T_\ix{c}$ is
independent of the contour index $j_1$: given $j_1 = -$,
$a^{j_1}(t_1)$ is sorted to the leftmost position of all operators
with $-$-index; given $j_1 = +$, $a^{j_1}(t_1)$ is sorted to the
rightmost position of all operators with $+$-index which is exactly
the same place in the total expression as for $j_1 = -$. As a
consequence
\begin{equation}
  \label{eq:causality1}
  G^{-, j_2 \ldots j_n|j'}(t|t') = G^{+, j_2 \ldots j_n|j'}(t|t'),
  \qquad
  \mbox{if}
  \quad
  t_1 > t_2, \ldots, t_n'.
\end{equation}
This property is at the heart of a theorem of causality which emerges
after transforming~\eref{eq:causality1} to the Keldysh basis. Let us
first transform only the contour index $j_1$ to the Keldysh basis:
\begin{eqnarray}
  \fl
  G^{\alpha_1,j_2\ldots j_n|j'}(t|t') 
  &= (D^{-1})^{\alpha_1|-}G^{-, j_2 \ldots j_n|j'}(t|t')
  + (D^{-1})^{\alpha_1|+}G^{+, j_2 \ldots j_n|j'}(t|t')
  \nonumber
  \\
  \fl
  &=
  \label{eq:causality}
  \big[(D^{-1})^{\alpha_1|-} + (D^{-1})^{\alpha_1|+} \big]
  G^{-, j_2 \ldots j_n|j'}(t|t'),
  \qquad \mbox{if}
  \quad
  t_1 > t_2, \ldots, t_n'.
\end{eqnarray}
For $\alpha_1 = 1$ we can use
\begin{equation}
  \label{eq:aa}
  (D^{-1})^{1|-} = \frac{1}{\sqrt{2}} = - (D^{-1})^{1|+}
\end{equation}
to see that~\eref{eq:causality} vanishes. After transforming
the remaining contour indices to the Keldysh basis, we find
\begin{equation}
  G^{1,\alpha_2\ldots \alpha_n|\alpha'}(t|t') = 0,
  \qquad
  \mbox{if}
  \quad
  t_1 > t_2 \ldots t_n'.
\end{equation}
The same line of argument holds if any other time from $t_1, \ldots,
t_n'$ is the maximum time. [If the maximum time is one of $t_1',
\ldots, t_n'$ we use $D^{-|1} = -D^{+|1}$ instead of~\eref{eq:aa}.]
This leads us to the following \emph{theorem of
  causality}~\cite{chou}: the Green function
$G^{\alpha|\alpha'}(t|t')$ vanishes if the Keldysh index associated
with the strictly greatest time argument equals~$1$.  (The theorem
does not apply when multiple time arguments have simultaneously the
maximum value. However, this particular configuration has a vanishing
measure in the $2n$-dimensional time space.)

A special case is given when all Keldysh indices equal~$1$,
\begin{equation}
  \label{eq:causality_relation_GF}
  G^{1 \ldots 1|1 \ldots 1} \equiv 0.
\end{equation}

If only one of all Keldysh indices equals $2$ the corresponding time
argument has to be the greatest one for a nonvanishing Green function.
This has a direct impact on the analytic properties of the Green
function in frequency space as we will show now.

Consider the Green function $G^{21\ldots 1|1\ldots 1}$ with all
Keldysh indices being $1$ except for $\alpha_1 = 2$. Due to the
theorem of causality $G^{21\ldots 1|1\ldots 1}(t_1 \ldots t_n|t'_1
\ldots t'_n)$ is nonvanishing only for $t_1 > t_2, \ldots, t_n'$. For
the Fourier transform~\eref{eq:freq_cons} this means that the
integration range of $t_2, \ldots t'_n$ is bounded from above by
$t_1$,
\begin{eqnarray}
  \fl
  \nonumber
  G^{21\ldots 1|1\ldots 1}(t_1 = 0, \omega_2, \ldots,
  \omega_n|\omega') =
  \int_{- \infty}^0 \! \rmd t_2 \ldots \rmd t'_n \,
  \rme^{\rmi(\omega_2 t_2 + \ldots + \omega_n t_n - \omega' \cdot t')}
  \\
  \fl
  \hspace{22em}
  \times
  G^{21\ldots 1|1\ldots 1}(t_1 = 0, t_2, \ldots, t_n|t').
  \label{eq:causality_FT}
\end{eqnarray}
$G^{21\ldots 1|1\ldots 1}(t|t')$ decays if any of $|t_k-t_1|,
|t'_k-t_1|$ goes to infinity [which makes the integral
in~\eref{eq:causality_FT} converge]. Due to the upper boundary for the
integration range of $t_2, \ldots, t_n$, only the decay for $t_2,
\ldots, t_n \rightarrow - \infty$ will create poles (branch-cuts) in
the upper half plane (uhp) of $\omega_2, \ldots \omega_n$. However,
the Green function will be analytic in the lower half plane (lhp) of
these frequencies since nonanalytic features in the lhp corresponding
to decays in the limit $t_2, \ldots, t_n \rightarrow +\infty$ do not
appear due to the restriction $t_2, \ldots, t_n < t_1$.  In the same
way, $G^{21\ldots 1|1\ldots 1}(t_1=0, \omega_2 \ldots
\omega_n|\omega'_1 \ldots \omega'_n)$ can be proven to be analytic in
the uhp of $\omega'_1 \ldots \omega'_n$. The differing half planes of
analyticity for the $\omega_k$ and $\omega_k'$ stem from the different
sign conventions for $\omega$ and $\omega'$ in the exponential
$\rme^{\rmi(\omega \cdot t - \omega' \cdot t')}$ of the Fourier transform.

The same line of argument holds if any other Keldysh index from
$\alpha_1, \ldots \alpha'_n$ equals $2$, the rest being $1$.  Hence we
can formulate a \emph{theorem of analyticity}: The Green functions
\begin{eqnarray}
    G^{21\ldots 1|1 \ldots 1}(t_1, \omega_2, \ldots, \omega_n|\omega'_1,
    \ldots, \omega'_n),
    \nonumber
    \\
    G^{121\ldots 1|1 \ldots 1}(\omega_1, t_2, \omega_3, \ldots,
    \omega_n|\omega'_1, \ldots, \omega'_n),
    \nonumber
    \\
    \ldots ,
    \\
    G^{1\ldots 1|1 \ldots 121}(\omega_1, \ldots,
    \omega_n|\omega'_1, \ldots, \omega'_{n-2}, t'_{n-1}, \omega'_n),
    \nonumber
    \\
    G^{1\ldots 1|1 \ldots 12}(\omega_1, \ldots,
    \omega_n|\omega'_1, \ldots, \omega'_{n-1}, t'_n)
    \nonumber
\end{eqnarray}
are analytic in the lhp of the frequencies $\omega_1, \ldots,
\omega_n$ and analytic in the uhp of the frequencies $\omega'_1,
\ldots, \omega'_n$. These $2n$ special Green functions describe higher
order response functions. Being formulated for bosons in time space,
they are correspondingly given by the expectation value of a series of
nested commutators~\cite{hao, chou}.

The irreducible vertex functions obey a \emph{theorem of causality} as
well: the vertex function $\gamma^{\alpha'|\alpha}(t'|t)$ vanishes if
the Keldysh index associated with the strictly greatest time argument
equals 2.

The proof of this theorem is based on the structure of the diagrams
contributing to $\gamma^{\alpha'|\alpha}(t'|t)$. Consider such a
one-particle irreducible diagram $d^{\alpha'|\alpha}(t'|t)$ and
suppose e.g. $\alpha_1 = 2$. We show that a nonvanishing contribution
to the diagram necessitates that at least one time argument from $t_2,
\ldots t'_n$ is greater or equal $t_1$.

The diagram is composed of bare two-particle vertices which are
connected by internal propagators. The vertices and propagators
contribute to the value of the diagram multiplicatively, as described
in~\eref{eq:amp_diagram}. Some of the vertices have one or
two amputated external lines which correspond to the indices of the
diagram.  When transformed to the Keldysh basis, the bare two-particle
vertex given in~\eref{eq:bare_vertex_time} acquires the form
\begin{equation}
  \fl
  \label{eq:bare_vertex_time_Keldysh}
  \bar v^{\alpha'_1 \alpha'_2|\alpha_1 \alpha_2}(t'_1, t'_2| t_1, t_2)
  \sim
  \cases{\delta(t_1 = t_2 = t'_1 = t'_2), &if $\alpha_1 +
    \alpha_2 + \alpha'_1 + \alpha'_2$ is odd,\\
    0, &else.}
\end{equation}
According to this equation all four time arguments of a given vertex
are equal. Therefore the vertex carrying $t_1$ as external time
argument has its other three time arguments equal to $t_1$ as well. If
there is an external one among those three, then $t_1$ is not the
strictly greatest external time argument. Let us now suppose that the
other three time arguments are internal ones. As $\alpha_1 = 2$, at
least one of the vertex' three internal Keldysh indices equals $1$,
see equation~\eref{eq:bare_vertex_time_Keldysh}. There is an internal
one-particle Green function attached to this Keldysh index $1$. At its
other end this Green function is attached to a Keldysh index $2$
(since $G^{1|1} = 0$) and a time $\tau_1 \ge t_1$; this follows from
the theorem of causality for the Green functions. Actually all four
time arguments of the vertex attached to the back-end of the Green
function are equal to $\tau_1 \ge t_1$. If there is an external one
among them, then $t_1$ is again not the strictly greatest external
time argument. If not, this vertex also has as least one Keldysh index
equal to~$1$. Again there is attached a one-particle Green function
leading to a vertex with time $\tau_2 \ge \tau_1 \ge t_1$. We can
proceed in this way until we arrive finally at a vertex with time
$\tau_m \ge \ldots \ge \tau_1 \ge t_1$ which carries an external time
argument. This external time argument $\tau_m \ge t_1$ concludes our
proof. (There is the possibility that we never end at a vertex with an
external line because we can get trapped in a loop of internal
vertices with identical time arguments; this case however is of
measure zero and vanishes when integrated over internal time
arguments.)

We mention two consequences of the theorem of causality for the vertex
function which can be derived in the same way as their counterparts
for the Green functions. First we conclude that
\begin{equation}
  \label{eq:causality_relation_vertex}
  \gamma^{2 \ldots 2|2 \ldots 2} \equiv 0.
\end{equation}
Second, we get a \emph{theorem of analyticity}, namely that the $2n$
vertex functions
\begin{eqnarray}
  \gamma^{12\ldots 2|2 \ldots 2}(t'_1, \omega'_2, \ldots,
  \omega'_n|\omega_1, \ldots, \omega_n), \nonumber
  \\
  \gamma^{212\ldots 2|2 \ldots 2}(\omega'_1, t'_2, \omega'_3, \ldots,
  \omega'_n|\omega_1, \ldots, \omega_n), \nonumber
  \\
  \ldots ,
  \\
  \gamma^{2\ldots 2|2 \ldots 212}(\omega'_1, \ldots, \omega'_n|\omega_1,
  \ldots, \omega_{n-2}, t_{n-1}, \omega_n), \nonumber
  \\
  \gamma^{2\ldots 2|2 \ldots 21}(\omega'_1, \ldots, \omega'_n|\omega_1,
  \ldots, \omega_{n-1}, t_n) \nonumber
\end{eqnarray}
are analytic in the lhp of the frequencies $\omega'_1, \ldots, \omega'_n$
and analytic in the uhp of the frequencies $\omega_1, \ldots, \omega_n$.

For the special case of one-particle functions, the causality
properties reduce to the well-known statements
\begin{equation}
  G^{1|1} \equiv 0, \qquad \Sigma^{2|2} \equiv 0,
\end{equation}
and $G^\ix{Ret}(\omega') \equiv G^{2|1}(t=0|\omega')$ is analytic in
the uhp of $\omega'$, $G^\ix{Av}(\omega) \equiv G^{1|2}(\omega|t'=0)$
is analytic in the lhp of $\omega$, $\Sigma^\ix{Ret}(\omega) \equiv
\Sigma^{1|2}(t'=0|\omega)$ is analytic in the uhp of $\omega$,
$\Sigma^\ix{Av}(\omega') \equiv \Sigma^{2|1}(\omega'|t=0)$ is analytic
in the lhp of $\omega'$.

%======================================================================

%====================================================================

\section{Equilibrium properties of the Green and vertex functions}
\label{sec:equilibrium}

We now focus on the case that the temperatures $T_k = 1/\beta_k$ and
chemical potentials $\mu_k$ which determine the reservoir density
matrices $\rho_\ix{res}^{(k)}$ in~\eref{eq:rho_res} are all equal,
\begin{equation}
  T_k = T,
  \qquad
  \mu_k = \mu,
  \qquad
  k = 1, \ldots , M.
\end{equation}
In the limits of infinite reservoir size and infinite transient time,
the system is then described by a global equilibrium distribution:
\begin{eqnarray}
  \nonumber
  \lo{G^{j|j'}_{q|q'}(t|t') =}
  (-\rmi)^n \Tr \rho(t_0) T_\ix{c} \, a_{q_1}^{(j_1)}(t_1) \ldots
  a_{q'_1}^{(j'_1) \dagger}(t'_1)
  \\
  \lo{\rightarrow}
  (-\rmi)^n \Tr \rho_H(t=0) T_\ix{c} \, a_{q_1}^{(j_1)}(t_1) \ldots
  a_{q'_1}^{(j'_1) \dagger}(t'_1)
  \quad
  \mbox{for}
  \quad
  L/u \gg -t_0 \rightarrow \infty
  \label{eq:equilibration}
\end{eqnarray}
with the static equilibrium density matrix
\begin{equation}
  \label{eq:eq_density_matrix}
  \rho_H(t=0) = \rho_H  = \frac{\rme^{- \beta(H - \mu N)}}{Z},
  \qquad
  Z = \Tr \rme^{- \beta(H - \mu N)}.
\end{equation}
Here $H$ is the total Hamiltonian~\eref{eq:full_H} including the
quantum system, reservoirs and coupling, and $N$ is the total particle
number, which is conserved, $[H, N] = 0$. A proof
of~\eref{eq:equilibration} for coordinate space Green functions has
been given recently in~\cite{doyon} for the particular case of a spin
impurity model.

In the present section we imply~\eref{eq:equilibration} to hold and
can therefore replace the expectation
value~\eref{eq:expectation_value} used in~\eref{eq:GF}
and~\eref{eq:tilde_GF} by
\begin{equation}
  \label{eq:expectation_value_eq}
  \langle A(t) \rangle
  =
  \Tr \rho_H A(t),
\end{equation}
where the Heisenberg picture now has the reference time $t=0$.

We just described how a thermal equilibrium situation is generated by
the coupling to large equilibrated reservoirs. In this way, equilibrium
and the stationary nonequilibrium introduced in
section~\ref{sec:system} are embedded in a common framework. However,
the following discussion holds also for equilibrium interacting bulk
systems. The only preconditions are the
equations~\eref{eq:eq_density_matrix}
and~\eref{eq:expectation_value_eq}, where $H$ is the total
Hamiltonian.

%--------------------------------------------------------------------

\subsection{Kubo-Martin-Schwinger conditions}
\label{subsec:KMS}

Any operator $A(t)$ in the Heisenberg picture fulfils
\begin{equation}
  A(t- \rmi \beta) 
  = \rme^{\beta H}A(t)\rme^{- \beta H}
  = \rho^{-1}_H \rme^{\beta \mu N} A(t) \rme^{- \beta \mu N} \rho_H.
\end{equation}
Using the cyclic property of the trace, we can reshuffle two operators
at times $t_A$, $t_B$ in a correlation function,
\begin{equation}
  \label{eq:KMS_simple}
  \Tr \rho_H A(t_A - \rmi \beta) B(t_B)
  =
  \Tr \rho_H B(t_B) \rme^{\beta \mu N} A(t_A) \rme^{- \beta \mu N}.
\end{equation}
We are going to apply this relation, also known as
Kubo-Martin-Schwinger (KMS) condition~\cite{kubo,martin}, to the
$n$-particle Green function. Given a set of contour indices $j = (j_1,
\ldots, j_n)$ and times $t = (t_1, \ldots, t_n)$ define $t - \rmi
\beta_+(j)$ as the set of times obtained from $t$ by adding $(- \rmi
\beta)$ to every $t_k$ with $j_k = +$. For example, given $n=4$ and $j
= (-,+,+,-)$, we have $t - \rmi \beta_+(j) = (t_1, t_2 - \rmi \beta,
t_3 - \rmi \beta, t_4)$. Let us prescribe that $T_\ix{c}$ orders
operators with complex time arguments according to the real parts of
their time arguments. Identifying $A$ in~\eref{eq:KMS_simple} with the
block of anti-time ordered operators from the $+$-branch and $B$ with
the block of time ordered operators from the $-$-branch, we find
\begin{equation}
  \fl
  \label{eq:KMS_time}
  G^{j|j'}_{q|q'}\Big(t - \rmi \beta_+(j)\Big|t' - \rmi
  \beta_+(j')\Big) 
  =
  \zeta^{M^{j|j'}} \rme^{\beta \mu m^{j|j'}} 
  \widetilde G^{\bar j| \bar{j'}}_{q|q'}(t|t')
  =
  \zeta^{m^{j|j'}} \rme^{\beta \mu m^{j|j'}} 
  \widetilde G^{\bar j| \bar{j'}}_{q|q'}(t|t')
\end{equation}
with $\widetilde G$ being defined in~\eref{eq:tilde_GF} and
$m^{j|j'}$ denoting the excess of creation over annihilation operators
on the $+$-branch,
\begin{equation}
  m^{j|j'}
  =
  \sum_{\substack{k = 1, \ldots, n:\cr j'_k = +}} 1 
  - \sum_{\substack{k = 1, \ldots, n:\cr  j_k = +}} 1.
\end{equation}
The factor $\zeta^{M^{j|j'}}$ in~\eref{eq:KMS_time} is
due to the different parity of the permutations which are required to
sort
\begin{equation}
  a^{(j_1)}(t_1) \ldots a^{(j_n)}(t_n)
  a^{(j'_n)\,\dagger}(t'_n) \ldots a^{(j'_1)\, \dagger}(t'_1)
\end{equation}
according to $T_\ix{c}$ and to sort
\begin{equation}
  a^{(\bar j_1)}(t_1) \ldots a^{(\bar j_n)}(t_n)
  a^{(\bar {j'_n})\,\dagger}(t'_n) \ldots a^{(\bar {j'_1})\, \dagger}(t'_1)
\end{equation}
according to $\widetilde T_{\ix{c}}$. This difference can be computed
from the number $M^{j|j'}$ of pairwise swaps of adjacent operators
needed to exchange the ordered block of operators from the $+$-branch
with the ordered block of operators from the $-$-branch,
\begin{equation}
  M^{j|j'}
  =
  \left[
    \sum_{k: j'_k = +} 1 + \sum_{k: j_k = +} 1
  \right]
  \left[
    \sum_{k: j'_k = -} 1 + \sum_{k: j_k = -} 1
  \right].
\end{equation}
In~\eref{eq:KMS_time} we have used that
\begin{equation}
  \zeta^{M^{j|j'}} 
  =
  \zeta^{m^{j|j'}},
\end{equation}
which is a consequence of having an even number $2n$ of indices.
The rewriting in terms of $m^{j|j'}$ turns out to be more convenient
in the following computation.

When transforming~\eref{eq:KMS_time} to Fourier space, we substitute on
the left hand side $t_k - \rmi \beta \rightarrow t_k$, $t'_k - \rmi
\beta \rightarrow t'_k$ for all time arguments on the $+$-branch and
find
\begin{eqnarray}
  \nonumber
  \fl
  \rme^{\beta \Delta^{j|j'}(\omega|\omega')}
  \int_{-\infty - \rmi \beta}^{\infty - \rmi \beta}
  \left[\prod_{k:j_k = +} \rmd t_k \right]
  \left[\prod_{k:j'_k = +} \rmd t'_k \right]
  \\
  \label{eq:KMS_freq_pre}
  \fl
  \hspace{5em}
  \times
  \int_{-\infty}^{\infty}
  \left[\prod_{k:j_k = -} \rmd t_k \right]
  \left[\prod_{k:j'_k = -} \rmd t'_k \right]
  \rme^{\rmi(\omega \cdot t - \omega' \cdot t')}
  G^{j|j'}_{q|q'}(t|t')
  =
  \zeta^{m^{j|j'}}
  \widetilde G^{\bar j| \bar{j'}}_{q|q'}(\omega|\omega'),
\end{eqnarray}
where $\Delta^{j|j'}(\omega|\omega')$ is defined as
\begin{equation}
  \Delta^{j|j'}(\omega|\omega')
  =
  \sum_{k: j'_k = +} (\omega'_k - \mu)
  - \sum_{k: j_k = +} (\omega_k - \mu).
\end{equation}
As next step we shift the path of integration for all time arguments
on the $+$-branch by $(+\rmi \beta)$ on the left hand side
of~\eref{eq:KMS_freq_pre} in order to obtain a standard Fourier
integral along the real axis.

\begin{figure}
  \centering
  \includegraphics{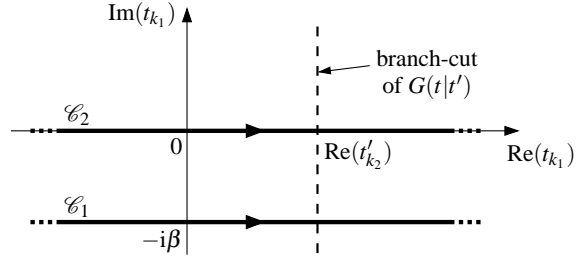}
  \caption{If an annihilation operator at time $t_{k_1}$ and a
    creation operator at time $t'_{k_2}$ are of the same state and are
    situated on the same branch of the contour, then the Green
    function $G(t|t')$ has a branch-cut along $\Real (t_{k_1}) = \Real
    (t'_{k_2})$. In this case the Fourier integrals of $G(t|t')$ over
    $t_{k_1}$ along the paths $\mathcal{C}_1$ and $\mathcal{C}_2$ are
    not equal.}
  \label{fig:branch_cut}
\end{figure}

This simultaneous shift of integration paths actually does not change
the value of the integral, as we explain now. In principle the shift
of a single time argument will change the value of the integral
because $G^{j|j'}(t|t')$ is in general not analytic in any single
$t_k$ or $t'_k$; branch-cuts occur due to changed time ordering
whenever a creation and annihilation operator of the same state and on
the same branch of the contour have coinciding real parts of their
time arguments. Suppose e.g. there exist $k_1, k_2 \in \{1, \ldots,
n\}$ with $j_{k_1} = j'_{k_2}=+$ and $q_{k_1} = q'_{k_2} =: q_0$, that
is an annihilation and a creation operator of the same state $q_0$ are
both situated on the $+$-branch. For real times $t_{k_1}, t'_{k_2}$
with $t_{k_1}$ approaching $t'_{k_2}$ from above, $t_{k_1} = t'_{k_2}
+ 0^+$, the contour ordered operator product appearing in~\eref{eq:GF}
will have the structure
\begin{equation}
  \label{eq:discont1}
  \ldots a^\dagger_{q_0}(t'_{k_2}) a_{q_0}(t'_{k_2}) \ldots,
\end{equation}
where the dots represent the other operators and a prefactor $\zeta$
if necessary. But for $t_{k_1}$ approaching $t'_{k_2}$ from below,
$t_{k_1} = t'_{k_2} - 0^+$, the structure is
\begin{equation}
  \label{eq:discont2}
  \zeta \ldots a_{q_0}(t'_{k_2}) a^\dagger_{q_0}(t'_{k_2}) \ldots
  =
  \zeta \ldots \big[1 + \zeta a^\dagger_{q_0}(t'_{k_2})
  a_{q_0}(t'_{k_2}) \big] \ldots.
\end{equation}
Here we use the commutation relation~\eref{eq:comm_rel}. The
results~\eref{eq:discont1} and~\eref{eq:discont2} differ by $\zeta$ so
that the Green function is discontinuous in $t_{k_1}$. When complex
time arguments are allowed for, the discontinuity is stretched to a
branch-cut in the complex $t_{k_1}$-plane as sketched in
figure~\ref{fig:branch_cut}.

However, we do not only shift the integration path of a single time
argument but those of all time arguments on the $+$-branch
simultaneously. If the dependence on the $n_+$ time arguments on the
$+$-branch is expressed through a centre time
\begin{equation}
  t_\ix{c} = \frac{1}{n_+} 
  \left[
    \sum_{k: j_k = +} t_k
    + \sum_{k: j'_k = +} t_k'
  \right]
\end{equation}
and $(n_+-1)$ relative times, then $G$ is analytic in $t_\ix{c}$: a
simultaneous change of all times on the $+$-branch by the same value
does not affect time ordering. Hence by shifting the integration path
of $t_\ix{c}$ while keeping all relative times unchanged we do not
alter the value of the integral. (Note that the dependence on the
relative time arguments might remain nonanalytic.) Therefore
equation~\eref{eq:KMS_freq_pre} is equivalent to
\begin{equation}
  \label{eq:KMS_freq}
  \rme^{\beta \Delta^{j|j'}(\omega|\omega')}
  G^{j|j'}_{q|q'}(\omega|\omega')
  =
  \zeta^{m^{j|j'}}
  \widetilde G^{\bar j| \bar{j'}}_{q|q'}(\omega|\omega').
\end{equation}

%----------------------------------------------------------------------

\subsection{Fluctuation dissipation theorem (FDT)}
\label{subsec:fluc_diss}

Equation~\eref{eq:KMS_freq} formulates a relation between $G$ and
$\widetilde G$. Since $\widetilde G$ is a rather artificial object of
no direct practical use, it is desirable to eliminate it
from~\eref{eq:KMS_freq} in favour of $G$.

In the one-particle case, we can make use of~\eref{eq:G_G_tilde} and
obtain
\begin{equation}
  \label{eq:fluct_diss_pre}
  \rme^{\beta \Delta^{j|j'}(\omega|\omega')} G^{j|j'}_{q|q'}(\omega|\omega')
  =
  \zeta^{m^{j|j'}} G^{j'|j}_{q|q'}(\omega|\omega'), \qquad n = 1.
\end{equation}
Plugging in $j,j' = \mp$ yields four equations, two of them being
tautologies and the other two being equivalent to
\begin{equation}
  \label{eq:fluct_diss_cont}
  G^<(\omega) = \zeta \rme^{-\beta(\omega - \mu)} G^>(\omega).
\end{equation}
The Green functions depend only on one frequency argument due to time
translational invariance.  A transformation to the Keldysh basis
yields the well-known FDT, which formulates a relation between the
correlation function $G^\ix{K}$ and the response functions
$G^\ix{Ret}$ and $G^\ix{Av}$,
\begin{equation}
  \label{eq:fluct_diss}
  G^\ix{K}(\omega) 
  =
  \left[1 + \zeta 2 n_\zeta (\omega-\mu) \right]
  \left[G^\ix{Ret}(\omega) -G^\ix{Av}(\omega) \right].
\end{equation}
Here
\begin{equation}
  \label{eq:n_zeta}
  n_\zeta (\omega) 
  = \frac{1}{\rme^{\beta \omega} - \zeta}
\end{equation}
is the Bose ($\zeta = +1$) or Fermi ($\zeta = -1$) function, and
\begin{equation}
  \label{eq:n_tanh_coth}
  1 + \zeta 2 n_\zeta (\omega)
  =
  \cases{\coth \frac{\beta \omega}{2}, &if $\zeta = +1$,
    \\ \tanh \frac{\beta \omega}{2}, &if $\zeta = -1$.}
\end{equation}

The FDT for the self-energy can be obtained by
combining~\eref{eq:fluct_diss}
with~\eref{eq:Dyson_Ret}~--~\eref{eq:Dyson_K},
\begin{equation}
  \Sigma^\ix{K}(\omega) 
  =
  \left[1 + \zeta 2 n_\zeta (\omega-\mu) \right]
  \left[\Sigma^\ix{Ret}(\omega) -\Sigma^\ix{Av}(\omega) \right].
\end{equation}
Transforming this result to the contour basis leads to an analogue
of~\eref{eq:fluct_diss_cont},
\begin{equation}
  \Sigma^<(\omega) = \zeta \rme^{-\beta(\omega - \mu)} \Sigma^>(\omega).
\end{equation}

%--------------------------------------------------------------------

\subsection{Time reversal}
\label{subsec:time_reversal}

A direct generalization of~\eref{eq:G_G_tilde} to the
multi-particle case does not exist. In order to eliminate $\widetilde
G$ from~\eref{eq:KMS_freq}, we should resort to time reversal.

The time reversal operator $\Theta$ is antiunitary, that is $\langle
\psi | \Theta \phi \rangle = \langle \phi | \Theta^\dagger \psi
\rangle$ and $\Theta \Theta^\dagger = \Theta^\dagger \Theta = 1$; for
details see e.g.~\cite{messiah}. Its action either on a
single-particle state $|q \rangle = |\bi{x}, m \rangle$ in the
eigenbasis of position and $s_z$ component of spin or on a state
$|\bi{p}, m \rangle$ in the eigenbasis of momentum and $s_z$ is given
by
\begin{eqnarray}
  \label{eq:xm_time_reversal}
  \Theta |\bi{x}, m \rangle 
  &=
  \rme^{\rmi \pi m} |\bi{x}, -m \rangle,
  \\
  \Theta |\bi{p}, m \rangle 
  &=
  \rme^{\rmi \pi m} | -\bi{p}, -m \rangle.
\end{eqnarray}
Since the vacuum state $|\mathrm{vac}\rangle$ is time reversal
invariant, $\Theta |\mathrm{vac}\rangle = |\mathrm{vac}\rangle$, the
transformation of creation and annihilation operators turns out to be
\begin{equation}
  \Theta a^\dagger_q \Theta^\dagger 
  =
  a^\dagger_{\Theta q},
  \qquad
  \Theta a_q \Theta^\dagger 
  =
  a_{\Theta q}.
\end{equation}
Explicitly, in the position or momentum basis it reads
\begin{eqnarray}
  \label{eq:psi_x_time_reversal}
  \Theta a^\dagger_{\bi{x}, m} \Theta^\dagger 
  =
  \rme^{\rmi \pi m} a^\dagger_{\bi{x}, -m},
  &
  \Theta a_{\bi{x}, m} \Theta^\dagger 
  =
  \rme^{-\rmi \pi m} a_{\bi{x}, -m},
  \\
  \label{eq:psi_p_time_reversal}
  \Theta a^\dagger_{\bi{p}, m} \Theta^\dagger 
  =
  \rme^{\rmi \pi m} a^\dagger_{-\bi{p}, -m},
  \qquad
  &
  \Theta a_{\bi{p}, m} \Theta^\dagger 
  =
  \rme^{-\rmi \pi m} a_{-\bi{p}, -m}.
\end{eqnarray}
We will designate a time reversed state $q$ or operator $A$ by a tilde,
\begin{equation}
  \widetilde q = \Theta q,
  \qquad
  \widetilde A = \Theta A \Theta^\dagger.
\end{equation}

If $\{ s \}$ is a complete set of orthonormal many-body states, so is
$\{ \Theta^\dagger s \}$. (Completeness of $\{ \Theta^\dagger s \}$
follows from
\begin{equation}
  \langle \psi \vert
  \left[ \sum_s \vert \Theta^\dagger s \rangle \langle \Theta^\dagger s
    \vert \right] \vert \phi \rangle
  =
  \sum_s \langle s \vert \Theta \psi \rangle \langle \Theta \phi \vert
  s \rangle
  =
  \langle \Theta \phi \vert \Theta \psi \rangle
  =
  \langle \psi \vert \phi \rangle,
\end{equation}
which holds for any $\psi, \phi$.) Hence, for any operator $A$
\begin{equation}
  \label{eq:trace_reverse1}
  \Tr A
  =
  \sum_s \langle \Theta^\dagger s| A | \Theta^\dagger s \rangle
  =
  \sum_s \langle s| \Theta A \Theta^\dagger | s \rangle^\ast
  =
  \left( \Tr \widetilde A \right)^\ast.
\end{equation}
The time reversed density matrix is given by
\begin{equation}
  \widetilde{\rho_H}
  =
  \Theta \frac{1}{Z} \rme^{-\beta (H -\mu N)} \Theta^\dagger
  =
  \frac{1}{Z} \rme^{-\beta (\widetilde H - \mu N)}
\end{equation}
with
\begin{equation}
  Z 
  =
  \Tr \rme^{-\beta (H -\mu N)}
  =
  \big( \Tr \Theta \rme^{-\beta (H -\mu N)} \Theta^\dagger \big)^\ast
  =
  \Tr \rme^{-\beta (\widetilde H -\mu  N)},
\end{equation}
where we used $N = \widetilde N$. Hence
\begin{equation}
  \label{eq:rho_time_rev}
  \widetilde{\rho_H}
  =
  \rho_{\widetilde H}.
\end{equation}

Applying time reversal to an operator in the Heisenberg picture yields
\begin{equation}
  \Theta A(t) \Theta^\dagger
  =
  \Theta \rme^{\rmi Ht} A \rme^{-\rmi Ht} \Theta^\dagger
  =
  \rme^{-\rmi \widetilde H t} \widetilde{A} \rme^{\rmi \widetilde H t}
  =
  \widetilde A(-t) \big|_{\widetilde H},
  \label{eq:heisenberg_time_rev}
\end{equation}
where the subscript $|_{\widetilde H}$ indicates that time evolution
is induced by $\widetilde H$ on the right hand side.

From~\eref{eq:trace_reverse1},~\eref{eq:rho_time_rev}
and~\eref{eq:heisenberg_time_rev} we conclude for the $n$-particle
tilde Green function defined in~\eref{eq:tilde_GF} that
\begin{eqnarray}
  \widetilde G^{j|j'}_{q|q'}(t|t') &= (-\rmi)^n \Tr \rho_H \widetilde
  T_\ix{c} \, a_{q_1}^{(j_1)}(t_1) \ldots a_{q'_1}^{(j'_1)\,
    \dagger}(t'_1) \nonumber
  \\
  &= (-\rmi)^n \Big[ \Tr \Theta \rho_H \widetilde T_\ix{c} \,
  a_{q_1}^{(j_1)}(t_1) \ldots a_{q'_1}^{(j'_1)\, \dagger}(t'_1)
  \Theta^\dagger \Big]^\ast \nonumber
  \\
  &= (-\rmi)^n \Big[ \Tr \rho_{\widetilde H} T_\ix{c} \, a_{\widetilde
    q_1}^{(j_1)}(-t_1)\big|_{\widetilde H} \ldots a_{\widetilde
    q_1'}^{(j'_1)\, \dagger}(-t'_1)\big|_{\widetilde H} \Big]^\ast
  \nonumber
  \\
  &= (-1)^n \left[G^{j|j'}_{\widetilde q|\widetilde
      q'}(-t|-t')\big|_{\widetilde H} \right]^\ast \nonumber
  \\
  &= G^{\bar {j'}|\bar j}_{\widetilde q'|\widetilde q}(-t'|-t)
  \big|_{\widetilde H}.
  \label{eq:GF_time_reversal}
\end{eqnarray}
Here $\widetilde q = (\widetilde q_1, \ldots, \widetilde q_n)$, and in
the last two lines the subscript $|_{\widetilde H}$ indicates that
both the time evolution and the equilibrium density matrix are
determined by $\widetilde H$. For the very last step, we use the rule
for complex conjugation formulated in~\eref{eq:GF(t)_conj}.  In
Fourier space, equation~\eref{eq:GF_time_reversal} reads
\begin{equation}
  \widetilde G^{j|j'}_{q|q'}(\omega|\omega')
  =
  G^{\bar {j'}|\bar j}_{\widetilde q'|\widetilde
    q}(\omega'|\omega) \big|_{\widetilde H}.
  \label{eq:GF_time_reversal_freq}
\end{equation}

%----------------------------------------------------------------------

\subsection{Combining KMS conditions and time reversal}
\label{subsec:gen_fluc_diss}

Unlike the property~\eref{eq:G_G_tilde} of $\widetilde G$ which is
restricted to the one-particle case,
equation~\eref{eq:GF_time_reversal_freq} holds for any $n$ and
provides a general possibility to eliminate $\widetilde G$
from~\eref{eq:KMS_freq}. Using it we find
\begin{equation}
  \label{eq:gen_fluc_diss}
  \rme^{\beta \Delta^{j|j'}(\omega|\omega')}
  G^{j|j'}_{q|q'}(\omega|\omega')
  =
  \zeta^{m^{j|j'}}
  G^{j'|j}_{\widetilde q'| \widetilde
    q}(\omega'|\omega) \big|_{\widetilde H},
\end{equation}
which is a generalization of~\eref{eq:fluct_diss_pre} to arbitrary
particle number $n$. We recover formula~\eref{eq:fluct_diss_pre}
from~\eref{eq:gen_fluc_diss} by means of
\begin{equation}
  \label{eq:1P_time_reversal}
  G^{j|j'}_{q|q'}(\omega|\omega')
  =
  G^{j|j'}_{\widetilde q'| \widetilde
    q}(\omega'|\omega)\big|_{\widetilde H} \; ,
  \quad
  n = 1,
\end{equation}
which follows from combining the Fourier transform
of~\eref{eq:G_G_tilde} with~\eref{eq:GF_time_reversal_freq}.

If one transforms equation~\eref{eq:gen_fluc_diss} to the Keldysh
basis, the result can be considered as a multi-particle FDT since the
components of the multi-particle Green functions in the Keldysh basis
have a distinct interpretation in terms of response and correlation
functions~\cite{hao}. We perform this transformation in
section~\ref{subsec:gen_fluc_diss_spec}, restricting ourselves to a
class of physically interesting systems which allow for certain
simplifications of the multi-particle FDT.

Before deriving an $n$-particle identity for the vertex function
analogous to~\eref{eq:gen_fluc_diss}, we note that the bare
two-particle interaction vertex given by~\eref{eq:bare_vertex}
satisfies
\begin{equation}
  \label{eq:fluc_diss_bare_vert}
  \rme^{\beta \Delta^{j'|j}(\omega'|\omega)}
  \bar v^{j'|j}_{q'|q}(\omega'|\omega)
  =
  \zeta^{m^{j'|j}}
  \bar v^{j|j'}_{\widetilde q| \widetilde q'}(\omega|\omega')
  \big|_{\widetilde H}.  
\end{equation}
In the case $(j'|j) \notin \{(--|--), (++|++)\}$ this equation is
automatically fulfilled because both quantities equal zero. For
$(j'|j) \in \{(--|--), (++|++)\}$
equation~\eref{eq:fluc_diss_bare_vert} states
\begin{equation}
  \bar v^{j'|j}_{q'|q}(\omega'|\omega)
  =
  \bar v^{j|j'}_{\widetilde q| \widetilde q'}(\omega|\omega')
  \big|_{\widetilde H}
  \quad
  \mbox{if}
  \quad
  j'_1 = j'_2 = j_1 = j_2 = \mp,
\end{equation}
which follows from 
\begin{eqnarray}
  \langle q'_1 q'_2| v \big(|q_1 q_2 \rangle +  \zeta |q_2 q_1 \rangle
  \big)
  &=
  \big(\langle q'_1 q'_2| + \zeta \langle q'_2 q'_1 | \big)
  \Theta^\dagger \Theta v \Theta^\dagger \Theta |q_1 q_2
  \rangle
  \nonumber
  \\
  &=
  \langle \widetilde q_1 \widetilde q_2| \widetilde v \big(|\widetilde
  q'_1 \widetilde q'_2 \rangle +  \zeta |\widetilde q'_2 \widetilde
  q'_1 \rangle \big),
\end{eqnarray}
where $\widetilde v = \Theta v \Theta^\dagger$.

With help of~\eref{eq:fluc_diss_bare_vert} we can now prove the
identity
\begin{eqnarray}
  \label{eq:gen_fluc_diss_vert}
  \rme^{\beta \Delta^{j'|j}(\omega'|\omega)}
  \gamma^{j'|j}_{q'|q}(\omega'|\omega)
  =
  \zeta^{m^{j'|j}} \gamma^{j|j'}_{\widetilde q| \widetilde q'}
  (\omega|\omega') \Big|_{\widetilde H}
\end{eqnarray}
by identifying one by one the diagrams contributing to each side of
it. The values of the individual diagrams are given
by~\eref{eq:amp_diagram}. The prefactors $\rme^{\beta
  \Delta^{j'|j}(\omega'|\omega)}$ and $\zeta^{m^{j'|j}}$ appearing
in~\eref{eq:gen_fluc_diss_vert} can be split and distributed among the
noninteracting single-particle Green functions and bare two-particle
vertices in~\eref{eq:amp_diagram} by making use of the properties
\begin{eqnarray}
  \label{eq:Delta_group}
  \Delta^{j'|j}(\omega'|\omega)
  =
  \Delta^{j'|i}(\omega'|\nu) + \Delta^{i|j}(\nu|\omega),
  \\
  \label{eq:m_group}
  m^{j'|j}
  =
  m^{j'|i} + m^{i|j},
\end{eqnarray}
which can be easily generalized to mixed particle numbers and chains
of arbitrary length, like e.g.
\begin{equation}
  m^{j'_1 j'_2|j_1 j_2}
  =
  m^{j'_1 j'_2|i_1 i_2} + m^{i_1|i'_1} + m^{i_2|i'_2} + m^{i'_1 i'_2|j_1
    j_2}.
\end{equation}
According to~\eref{eq:gen_fluc_diss} and~\eref{eq:fluc_diss_bare_vert}
the constituents of both sides of~\eref{eq:gen_fluc_diss_vert} can
then be readily identified.

%----------------------------------------------------------------------

\subsection{Systems with special behaviour under time
  reversal}
\label{subsec:time_rev_inv_GF}

For a further evaluation of~\eref{eq:gen_fluc_diss}
and~\eref{eq:gen_fluc_diss_vert}, we should establish how the system
behaves under time reversal. We restrict the following discussion to
the class of systems which satisfy
\begin{equation}
  \label{eq:time_rev_inv_GF}
  G^{j|j'}_{q|q'}(t|t')
  =
  G^{j|j'}_{\widetilde q| \widetilde q'}(t|t') \big|_{\widetilde H}
\end{equation}
for all indices and time arguments. Note that
equation~\eref{eq:time_rev_inv_GF} is not only a condition on the
Hamiltonian but also on the basis of single-particle states.

We mention some examples which demonstrate the extent of this class of
systems. A trivial example for the validity
of~\eref{eq:time_rev_inv_GF} is given when the Hamiltonian and the
single-particle states are time reversal invariant.

As a less trivial example we consider the single impurity Anderson model (SIAM)
 \cite{anderson} in a magnetic field $B$. This model consists of an electronic
impurity with on-site interaction which is coupled to a conductor. The 
corresponding Hamiltonian reads
\begin{eqnarray}
  \nonumber
  H_B
  = 
  \sum_\sigma (\epsilon_0 + \sigma B) a^\dagger_\sigma a_\sigma
  +
  \sum_\sigma \int d^3 p \; \epsilon_p c^\dagger_{\bi{p}, \sigma}
  c_{\bi{p},\sigma}
  \\
  \label{eq:H_B}
  \hspace{5em}
  + \sum_\sigma \int d^3 p \; \big[V_p a^\dagger_\sigma c_{\bi{p},\sigma}
  + \mbox{H.c.}\big]  
  +
  U a^\dagger_\uparrow a_\uparrow a^\dagger_\downarrow a_\downarrow.
\end{eqnarray}
Here $\sigma = \pm \frac{1}{2} = \uparrow, \downarrow$ denote the
eigenvalues of the single-particle spin $s_z$, and $c^\dagger, c$ are
creators and annihilators of states in the conductor. 

As a single-particle basis to be used in~\eref{eq:time_rev_inv_GF} we
choose the common eigenbasis of position and $s_z$, $|q\rangle = |
\bi{x}, \sigma \rangle$, where the values of $\bi{x}$ are considered
to label the states of both the conductor and the impurity.

In order to verify~\eref{eq:time_rev_inv_GF} for this Hamiltonian and
this choice of basis, we make use of the unitary transformation
$\Omega = \rme^{-\rmi \pi S_x}$ rotating spin space by $\pi$ around the
$x$-axis, $S_x$ being the $x$-component of the total spin,
\begin{equation}
  \label{eq:xm_spin_rot}
  \Omega | \bi{x}, \sigma \rangle
  = 
  -\rmi | \bi{x}, -\sigma \rangle.
\end{equation}
Using~\eref{eq:psi_p_time_reversal} and the fact that $\epsilon_p$ and
$V_p$ do not depend on the sign of $\bi{p}$ in~\eref{eq:H_B}, we find
\begin{equation}
  \label{eq:H_B_time_reversal}
  \widetilde{H_B} = H_{-B} = \Omega H_B \Omega^\dagger.
\end{equation}
Since a combination of~\eref{eq:xm_time_reversal}
and~\eref{eq:xm_spin_rot} yields additionally
\begin{equation}
  | \widetilde q \rangle
  =
  \Theta | \bi{x}, \sigma \rangle
  =
  \rme^{\rmi \pi (\sigma + \frac{1}{2})} \Omega | \bi{x}, \sigma \rangle
  =
  \rme^{\rmi \pi (\sigma + \frac{1}{2})} \Omega | q\rangle
\end{equation}
we conclude
\begin{equation}
  G_{\widetilde q|\widetilde q'} \big|_{\widetilde H}
  =
  \rme^{\rmi \pi \sum_k (\sigma_k' - \sigma_k)}
  G_{\Omega q| \Omega q'} \big|_{\Omega H \Omega^\dagger}
  =
  G_{\Omega q| \Omega q'} \big|_{\Omega H \Omega^\dagger}
  \label{eq:Theta_Omega}
\end{equation}
with $\Omega q = (\Omega q_1, \ldots, \Omega q_n)$.
In~\eref{eq:Theta_Omega} we exploit that the Green function is
nonvanishing only for
\begin{equation}
  \sum_{k=1}^N \sigma'_k
  =
  \sum_{k=1}^N \sigma_k
\end{equation}
due to spin conservation. It is straightforward to derive
from~\eref{eq:GF},~\eref{eq:eq_density_matrix}
and~\eref{eq:expectation_value_eq} that
\begin{equation}
  \label{eq:G_unitary_transf}
  G_{\Omega q|\Omega q'}\big|_{\Omega H \Omega^\dagger}
  =
  G_{q|q'}.
\end{equation}
[Note that~\eref{eq:G_unitary_transf} holds in general for any unitary
operator $\Omega$.] Combining~\eref{eq:Theta_Omega}
and~\eref{eq:G_unitary_transf} we
conclude that the condition~\eref{eq:time_rev_inv_GF} is fulfilled for
the impurity model~\eref{eq:H_B} and the single-particle basis
$|q\rangle = | \bi{x}, \sigma \rangle$.

On the full analogy it can be shown that~\eref{eq:time_rev_inv_GF} is
also valid for this impurity model and the single-particle basis
$|q\rangle = | \bi{p}, \sigma \rangle$.
In~\eref{eq:H_B_time_reversal}~--~\eref{eq:G_unitary_transf}, $\Omega$
simply has to be replaced by $\Omega \Pi$, where $\Pi$ is the parity
transformation,
\begin{equation}
  \Pi | \bi{p}, \sigma \rangle
  = 
  | -\bi{p}, \sigma \rangle.
\end{equation}

As an example of a system which does not
fulfil~\eref{eq:time_rev_inv_GF} we consider the Hamiltonian
\begin{equation}
  H_{\bi{A}} 
  = 
  \int \rmd^3 x \; a^\dagger_{\bi{x}} \frac{\big[- \rmi \nabla - e
      \bi{A}(\bi{x})\big]^2}{2m} a_{\bi{x}}
\end{equation}
for spinless particles with charge $e$ moving in the field of a vector
potential $\bi{A}(\bi{x})$. Let us choose the position eigenstates as
the single-particle basis. Since
\begin{equation}
  \widetilde{H_{\bi{A}}}
  =
  H_{- \bi{A}},
\end{equation}
we conclude that for a generic vector potential
\begin{equation}
  G_{\widetilde{\bi{x}_1} \ldots
    \widetilde{\bi{x}_n}|\widetilde{\bi{x}'_1}
    \ldots \widetilde{\bi{x}'_n} }
  \Big|_{\widetilde{H_{\bi{A}}}}
  =
  G_{\bi{x}_1 \ldots \bi{x}_n|\bi{x}'_1 \ldots \bi{x}'_n }
  \Big|_{H_{-\bi{A}}}
  \neq
  G_{\bi{x}_1 \ldots \bi{x}_n|\bi{x}'_1 \ldots \bi{x}'_n }
  \Big|_{H_{\bi{A}}}.
\end{equation}
Obviously, equation~\eref{eq:time_rev_inv_GF} does not hold for this
choice of Hamiltonian and single-particle basis.

%----------------------------------------------------------------------

\subsection{Generalized fluctuation dissipation theorem for systems
  satisfying equation~\eref{eq:time_rev_inv_GF}}

\label{subsec:gen_fluc_diss_spec}

When relation~\eref{eq:time_rev_inv_GF} holds,
equation~\eref{eq:gen_fluc_diss} takes the form
\begin{equation}
  \label{eq:gen_fluc_diss_spec}
  \rme^{\beta \Delta^{j|j'}(\omega|\omega')}
  G^{j|j'}_{q|q'}(\omega|\omega')
  =
  \zeta^{m^{j|j'}} 
  G^{j'|j}_{q'|q}(\omega'|\omega).
\end{equation}
Additionally, equation~\eref{eq:time_rev_inv_GF} can be expanded in
orders of the two-particle interaction. From the first-order
term of this expansion of the two-particle Green function, we
infer that the bare two-particle interaction vertex complies with
\begin{equation}
  \bar v^{j'|j}_{q'|q}(\omega'|\omega)
  =
  \bar v^{j'|j}_{\widetilde q'|\widetilde
    q}(\omega'|\omega)\big|_{\widetilde H}\;.
\end{equation}
When combined with~\eref{eq:fluc_diss_bare_vert} it brings about the
identity
\begin{equation}
  \label{eq:gen_fluc_diss_spec_bare_vert}
  \rme^{\beta \Delta^{j'|j}(\omega'|\omega)}
  \bar v^{j'|j}_{q'|q}(\omega'|\omega)
  =
  \zeta^{m^{j'|j}}
  \bar v^{j|j'}_{q| q'}(\omega|\omega').
\end{equation}
Similarly to the derivation of~\eref{eq:gen_fluc_diss_vert} we deduce
from~\eref{eq:gen_fluc_diss_spec}
and~\eref{eq:gen_fluc_diss_spec_bare_vert} that
\begin{equation}
  \label{eq:gen_fluc_diss_spec_vert}
  \rme^{\beta \Delta^{j'|j}(\omega'|\omega)}
  \gamma^{j'|j}_{q'|q}(\omega'|\omega)
  =
  \zeta^{m^{j'|j}}
  \gamma^{j|j'}_{q|q'}(\omega|\omega').
\end{equation}
It is convenient to combine~\eref{eq:gen_fluc_diss_spec}
and~\eref{eq:gen_fluc_diss_spec_vert} with~\eref{eq:GF_conj_cont}
and~\eref{eq:vert_conj_cont}, respectively, in order to obtain the
equations
\begin{eqnarray}
  \label{eq:gen_fluc_diss_special}
  \rme^{\beta \Delta^{j|j'}(\omega|\omega')}
  G^{j|j'}_{q|q'}(\omega|\omega')
  &=
  (-1)^n \zeta^{m^{j|j'}} 
  G^{\bar j|\bar {j'}}_{q|q'}(\omega|\omega')^\ast,
  \\
  \label{eq:gen_fluc_diss_special_vert}
  \rme^{\beta \Delta^{j'|j}(\omega'|\omega)}
  \gamma^{j'|j}_{q'|q}(\omega'|\omega)
  &=
  - \zeta^{m^{j'|j}}
  \gamma^{\bar {j'}|\bar j}_{q'|q}(\omega'|\omega)^\ast,
\end{eqnarray}
where state indices and frequency arguments are now identical on both
sides.

We are going to transform~\eref{eq:gen_fluc_diss_special}
and~\eref{eq:gen_fluc_diss_special_vert} to the Keldysh basis in order
to achieve a multi-particle FDT. In what follows the state indices are
omitted for brevity. According to
\begin{equation}
  G^{j|j'} 
  = 
  \sum_{\alpha, \alpha'} D^{j|\alpha} G^{\alpha|\alpha'}
  (D^{-1})^{\alpha'|j'} 
\end{equation}
every component of $G^{j|j'}$ is expressed as a certain linear
combination of components of $G^{\alpha|\alpha'}$. Let us define
\begin{equation}
  \label{eq:G_epsilon}
  G^{j|j'}_\epsilon 
  = 
  \sum_{\substack{\alpha, \alpha'\cr (-1)^{\sum_k (\alpha_k +
        \alpha'_k)} = \epsilon}}
  D^{j|\alpha} G^{\alpha|\alpha'} (D^{-1})^{\alpha'|j'}
\end{equation}
for $\epsilon = \pm 1$ (cf.~\cite{chou, wang}).  It is obvious that
\begin{equation}
  \label{eq:G+-}
  G^{j|j'}  
  =
  G^{j|j'}_+ + G^{j|j'}_-.
\end{equation}
Using the relations
\begin{eqnarray}
  D^{j|\alpha} 
  &=
  (-1)^{\sum_k \alpha_k} D^{\bar j | \alpha},
  \\
  (D^{-1})^{\alpha|j}
  &=
  (-1)^{\sum_k \alpha_k} (D^{-1})^{\alpha|\bar j},
\end{eqnarray}
which follow from~\eref{eq:D}~--~\eref{eq:D_inv_single}, we find
additionally that
\begin{equation}
  \label{eq:Gbar+-}
  G^{\bar j|\bar{j'}}  
  =
  G^{j|j'}_+ - G^{j|j'}_-.
\end{equation}
Plugging~\eref{eq:G+-} and~\eref{eq:Gbar+-}
into~\eref{eq:gen_fluc_diss_special} and splitting the latter into the
real and the imaginary parts we obtain
\begin{eqnarray}
  \fl
  \label{eq:fluct_diss_gen_Re1}
  \sinh \case{\beta \Delta^{j|j'}(\omega|\omega')}{2} \;
  \Real G^{j|j'}_{\zeta \epsilon_n^{j|j'}}(\omega|\omega')
  =
  - \cosh \case{\beta \Delta^{j|j'}(\omega|\omega')}{2} \;
  \Real G^{j|j'}_{- \zeta \epsilon_n^{j|j'}}(\omega|\omega'),
  \\
  \fl
  \label{eq:fluct_diss_gen_Im1}
  \cosh \case{\beta \Delta^{j|j'}(\omega|\omega')}{2} \;
  \Imag G^{j|j'}_{\zeta \epsilon_n^{j|j'}}(\omega|\omega')
  =
  - \sinh \case{\beta \Delta^{j|j'}(\omega|\omega')}{2} \;
  \Imag G^{j|j'}_{- \zeta \epsilon_n^{j|j'}}(\omega|\omega'),
\end{eqnarray}
where
\begin{equation}
  \label{eq:epsilon}
  \epsilon_n^{j|j'} 
  = 
  (-1)^n \zeta^{1 + m^{j|j'}}.
\end{equation}
Utilizing~\eref{eq:n_tanh_coth} we finally achieve the representation
\begin{eqnarray}
  \label{eq:fluct_diss_gen_Re}
  \Real G^{j|j'}_{\epsilon_n^{j|j'}}(\omega|\omega')
  =
  - \Big[1 + \zeta 2 n_\zeta\big(\Delta^{j|j'}(\omega|\omega')\big)
  \Big] 
  \Real G^{j|j'}_{- \epsilon_n^{j|j'}}(\omega|\omega'),
  \\
  \label{eq:fluct_diss_gen_Im}
  \Imag G^{j|j'}_{- \epsilon_n^{j|j'}}(\omega|\omega')
  =
  - \Big[1 + \zeta 2 n_\zeta\big(\Delta^{j|j'}(\omega|\omega')\big)
  \Big] 
\Imag G^{j|j'}_{ \epsilon_n^{j|j'}}(\omega|\omega').
\end{eqnarray}
These two equations express relations between certain linear
combinations of the components of $G^{\alpha|\alpha'}$. Since the
components of $G^{\alpha|\alpha'}$ can be understood in terms of
higher order response and correlation functions~\cite{hao},
equations~\eref{eq:fluct_diss_gen_Re} and~\eref{eq:fluct_diss_gen_Im}
constitute a multi-particle FDT. Both equations can be evaluated for
$2^{2n}$ different realizations of the indices $(j|j')$. However, the
underlying equation~\eref{eq:gen_fluc_diss_special} is equivalent for
$G^{j|j'}$ and for $G^{\bar j|\bar{j'}}$ as can be inferred from
$m^{j|j'} = -m^{\bar j|\bar {j'}}$ and $\Delta^{j|j'}(\omega|\omega')
= -\Delta^{\bar j|\bar{j'}}(\omega|\omega')$. Therefore only half of
the $2^{2n}$ equations~\eref{eq:fluct_diss_gen_Re} are independent.
The same applies to~\eref{eq:fluct_diss_gen_Im}.

For bosons the function
$n_{\zeta=+}\big(\Delta^{j|j'}(\omega|\omega')\big)$ diverges in the
special case $\Delta^{j|j'}(\omega|\omega') \equiv 0$ (which occurs
for all contour indices being $-$ or all being $+$), and one then
should rather use~\eref{eq:fluct_diss_gen_Re1}
and~\eref{eq:fluct_diss_gen_Im1} with $\cosh [\beta
\Delta^{j|j'}(\omega|\omega')/2] = 1$ and $\sinh [\beta
\Delta^{j|j'}(\omega|\omega')/2] = 0$ instead
of~\eref{eq:fluct_diss_gen_Re} and~\eref{eq:fluct_diss_gen_Im}.

The transformation of~\eref{eq:gen_fluc_diss_special_vert} to the
Keldysh basis is completely analogous to that
of~\eref{eq:gen_fluc_diss_special}. Defining for $\epsilon = \pm 1$
\begin{equation}
  \gamma^{j'|j}_\epsilon 
  = 
  \sum_{\substack{\alpha', \alpha\cr (-1)^{\sum_k (\alpha'_k +
        \alpha_k)} = \epsilon}}
  D^{j'|\alpha'} \gamma^{\alpha'|\alpha} (D^{-1})^{\alpha|j}
\end{equation}
we end up with
\begin{eqnarray}
  \label{eq:fluct_diss_gen_vert_Re}
  \Real \gamma^{j'|j}_{\epsilon_1^{j'|j}}(\omega'|\omega)
  =
  - \Big[1 + \zeta 2 n_\zeta\big(\Delta^{j'|j}(\omega'|\omega)\big)
  \Big] 
  \Real \gamma^{j'|j}_{- \epsilon_1^{j'|j}}(\omega'|\omega),
  \\
  \label{eq:fluct_diss_gen_vert_Im}
  \Imag \gamma^{j'|j}_{- \epsilon_1^{j'|j}}(\omega'|\omega)
  =
  - \Big[1 + \zeta 2 n_\zeta\big(\Delta^{j'|j}(\omega'|\omega)\big)
  \Big] 
  \Imag \gamma^{j'|j}_{ \epsilon_1^{j'|j}}(\omega'|\omega),
\end{eqnarray}
where
\begin{equation}
  \epsilon_1^{j'|j} 
  = 
  - \zeta^{1 + m^{j'|j}}
\end{equation}
[cf.~\eref{eq:epsilon}]. Like in the case of Green functions, each of
equations~\eref{eq:fluct_diss_gen_vert_Re}
and~\eref{eq:fluct_diss_gen_vert_Im} contains only $2^{2n-1}$
independent relations.

Let us illustrate
equations~\eref{eq:fluct_diss_gen_Re},~\eref{eq:fluct_diss_gen_Im}
and~\eref{eq:fluct_diss_gen_vert_Re},~\eref{eq:fluct_diss_gen_vert_Im}
by some examples.  In the one-particle case, $n=1$, we
evaluate~\eref{eq:fluct_diss_gen_Re} and~\eref{eq:fluct_diss_gen_Im}
for $(j|j')=(-|-)$ and $(j|j')=(-|+)$. Since $G^{-|-} =
\frac{1}{2}(G^{1|2} + G^{2|1} + G^{2|2})$ and $G^{-|+} =
\frac{1}{2}(G^{1|2} - G^{2|1} + G^{2|2})$, we have
\begin{eqnarray}
  G^{-|-}_+
  &= \frac{1}{2} G^{2|2}
  =
  \frac{1}{2} G^\ix{K},
  \\
  G^{-|-}_-
  &= \frac{1}{2} (G^{1|2} + G^{2|1})
  = \frac{1}{2} (G^\ix{Av} + G^\ix{Ret}),
  \\
  G^{-|+}_+
  &= \frac{1}{2} G^{2|2}
  =
  \frac{1}{2} G^\ix{K},
  \\
  G^{-|+}_-
  &= \frac{1}{2} (G^{1|2} - G^{2|1})
  = \frac{1}{2} (G^\ix{Av} - G^\ix{Ret}).
\end{eqnarray}
Furthermore, we use
\begin{eqnarray}
  \Delta^{-|-}(\omega|\omega')
  =0,
  &\epsilon_1^{-|-}
  = - \zeta,
  \\
  \Delta^{-|+}(\omega|\omega')
  =\omega'-\mu,
  \qquad
  &\epsilon_1^{-|+}
  = - 1,
\end{eqnarray}
and the representation of the one-particle Green function
$G(\omega|\omega') = 2 \pi \delta(\omega-\omega')G(\omega)$. Then
equations~\eref{eq:fluct_diss_gen_Re} and~\eref{eq:fluct_diss_gen_Im}
evaluated for $(j|j')=(-|-), (-|+)$ read
\begin{eqnarray}
  \label{eq:fluc_diss_gen_1p_1}
  \Real G^\ix{K}(\omega)
  = 0,
  \\
  \Real \left[G^\ix{Av} (\omega) - G^\ix{Ret} (\omega)\right]
  =
  - \left[1 + \zeta 2 n_\zeta (\omega-\mu) \right]
  \Real G^\ix{K}(\omega) 
  = 0,
  \\
  \Imag \left[G^\ix{Av} (\omega) + G^\ix{Ret} (\omega) \right]
  = 0,
  \\
  \Imag G^\ix{K} (\omega)
  =
  - \left[1 + \zeta 2 n_\zeta (\omega-\mu) \right]
  \Imag \left[G^\ix{Av} (\omega) - G^\ix{Ret} (\omega) \right].
  \label{eq:fluc_diss_gen_1p_4}
\end{eqnarray}
These formulae obviously reproduce~\eref{eq:GF_conj_1p}
and~\eref{eq:fluct_diss} under the additional constraint
$G_{q|q'}(\omega) = G_{q'|q}(\omega)$ which follows from the special
one-particle property~\eref{eq:1P_time_reversal} and the
assumption~\eref{eq:time_rev_inv_GF}.

Evaluating~\eref{eq:fluct_diss_gen_vert_Re}
and~\eref{eq:fluct_diss_gen_vert_Im} for the one-particle vertex (the
self-energy), leads to a result similar
to~\eref{eq:fluc_diss_gen_1p_1}~--~\eref{eq:fluc_diss_gen_1p_4}: one
only needs to replace $G$ therein by the self-energy $\Sigma$.

For $n=2$ the complete set of relations for the Green functions is
obtained by evaluating~\eref{eq:fluct_diss_gen_Re}
and~\eref{eq:fluct_diss_gen_Im} for eight independent choices for
$(j_1 j_2|j'_1 j'_2)$ from 16 existing possibilities. Let us for
example discuss the three particular combinations $(j_1 j_2|j'_1 j'_2)
= (--|--), (-+|--), (--|++)$. Since
\begin{eqnarray}
  \Delta^{--|--}(\omega_1, \omega_2|\omega'_1, \omega'_2)
  =0,
  &
  \epsilon_2^{--|--}
  = \zeta,
  \\
  \Delta^{-+|--}(\omega_1, \omega_2|\omega'_1, \omega'_2)
  =-\omega_2 + \mu,
  &
  \epsilon_2^{-+|--}
  = +1,
  \\
  \Delta^{--|++}(\omega_1, \omega_2|\omega'_1, \omega'_2)
  =\omega'_1 + \omega'_2 - 2 \mu,
  \qquad
  &
  \epsilon_2^{--|++}
  = \zeta,
\end{eqnarray}
equations~\eref{eq:fluct_diss_gen_Re}
and~\eref{eq:fluct_diss_gen_Im} read
\begin{eqnarray}
  \Real G^{--|--}_-(\omega_1,\omega_2|\omega'_1,\omega'_2) 
  = 0,
  \\
  \Imag G^{--|--}_+(\omega_1,\omega_2|\omega'_1,\omega'_2) 
  = 0,
\end{eqnarray}
\begin{eqnarray}
  \fl
  \Real G^{-+|--}_+(\omega_1,\omega_2|\omega'_1,\omega'_2) 
  =
  - \big[1 + \zeta 2 n_\zeta(-\omega_2 + \mu) \big] 
  \Real G^{-+|--}_-(\omega_1,\omega_2|\omega'_1,\omega'_2), 
  \\
  \fl
  \Imag G^{-+|--}_-(\omega_1,\omega_2|\omega'_1,\omega'_2) 
  =
  - \big[1 + \zeta 2 n_\zeta(-\omega_2 + \mu) \big] 
  \Imag G^{-+|--}_+(\omega_1,\omega_2|\omega'_1,\omega'_2), 
\end{eqnarray}
\begin{eqnarray}
  \fl
  \Real G^{--|++}_\zeta(\omega_1,\omega_2|\omega'_1,\omega'_2) 
  =
  - \big[1 + \zeta 2 n_\zeta(\omega'_1 + \omega'_2 - 2 \mu) \big] 
  \Real G^{--|++}_{-\zeta}(\omega_1,\omega_2|\omega'_1,\omega'_2), 
  \\
  \fl
  \Imag G^{--|++}_{-\zeta}(\omega_1,\omega_2|\omega'_1,\omega'_2) 
  =
  - \big[1 + \zeta 2 n_\zeta(\omega'_1 + \omega'_2 - 2 \mu) \big] 
  \Imag G^{--|++}_\zeta(\omega_1,\omega_2|\omega'_1,\omega'_2). 
\end{eqnarray}
The functions $G^{j_1j_2|j'_1j'_2}_\pm$ appearing in these equations
can be found by explicit evaluation of~\eref{eq:G_epsilon} with help
of~\eref{eq:D}~--~\eref{eq:D_inv_single}. This results in
\begin{eqnarray}
  \fl
  G^{--|--}_+
  =
  \frac{1}{4} \big( G^{11|11} + G^{11|22} + G^{12|12} + G^{12|21}
  + G^{21|12} + G^{21|21} + G^{22|11} \big),
  \\
  \fl
  G^{--|--}_-
  =
  \frac{1}{4} \big( G^{11|12} + G^{11|21} + G^{12|11} + G^{21|11}
  + G^{22|21} + G^{22|12} + G^{21|22} + G^{12|22} \big),
\end{eqnarray}
\begin{eqnarray}
  \fl
  G^{-+|--}_+
  =
  \frac{1}{4} \big(- G^{11|11} - G^{11|22} + G^{12|12} + G^{12|21}
  - G^{21|12} - G^{21|21} + G^{22|11} \big),
  \\
  \fl
  G^{-+|--}_-
  =
  \frac{1}{4} \big(- G^{11|12} - G^{11|21} + G^{12|11} - G^{21|11}
  + G^{22|21} + G^{22|12} - G^{21|22} + G^{12|22} \big),
\end{eqnarray}
\begin{eqnarray}
  \fl
  G^{--|++}_+
  =
  \frac{1}{4} \big( G^{11|11} + G^{11|22} - G^{12|12} - G^{12|21}
  - G^{21|12} - G^{21|21} + G^{22|11} \big),
  \\
  \fl
  G^{--|++}_-
  =
  \frac{1}{4} \big(-G^{11|12} - G^{11|21} + G^{12|11} + G^{21|11}
  - G^{22|21} - G^{22|12} + G^{21|22} + G^{12|22} \big).
\end{eqnarray}

%====================================================================

%====================================================================

\section{Conservation of the properties in approximations}
\label{sec:approx}

In practice the Green and vertex functions can often be computed only
in approximations. In this section, we address the question whether
such approximations are consistent with the exact relations presented
in this review. Since the answer depends on the details of the
corresponding method, we have to restrict ourselves to specific
approximation schemes. First we discuss the large class of
diagrammatic approximations, and then we analyse approximations within
the framework of the fRG.

\subsection{Diagrammatic approximations}
\label{subsec:diagr}

Many of the approximations used to compute the Green or vertex
functions are given by taking into account only a subset of all
diagrams contributing to that function. This may be either a finite
set like in perturbation theory or an infinite one like in the random
phase approximation. We call this approach a diagrammatic
approximation and assume that apart from choosing only a subset of
diagrams no further approximation is done. Since all discussed
relations are linear in the vertex functions it follows that if two
distinct diagrammatic approximations satisfy a given relation each, so
does their sum.

\begin{figure}
  \centering
  \includegraphics{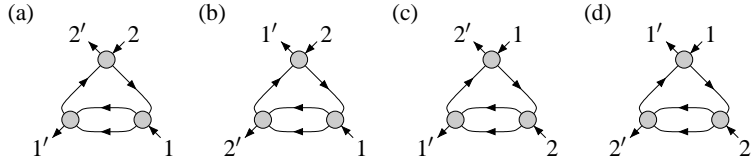}
  \caption{Examples of diagrams contributing to the two-particle
    vertex function $\gamma_{1'2'|12}$. The external indices represent
    contour index, state and frequency, e.g. $1 = (j_1, q_1,
    \omega_1)$.}
    \label{fig:approx}
\end{figure}

Consider a diagrammatic approximation to a multi-particle Green or
vertex function. The condition for the approximation to fulfil the
relation~\eref{eq:G_permute} or~\eref{eq:gamma_permute} under
permutations of particles is that the set of diagrams taken into
account is invariant under those permutations. For instance neither of
the diagrams in figure~\ref{fig:approx} alone fulfils the
relation~\eref{eq:gamma_permute}. The approximation given by the sum
of diagram (a) and (b) in figure~\ref{fig:approx} satisfies a part
of~\eref{eq:gamma_permute}, namely
\begin{equation}
  \gamma^{Pj'|j}_{Pq'|q}(P\omega'|\omega)
  =
  \zeta^P \gamma^{j'|j}_{q'|q}(\omega'|\omega).
\end{equation}
The same is true for the sum of (c) and (d). The sum of (a) and (c)
satisfies
\begin{equation}
  \gamma^{j'|Pj}_{q'|Pq}(\omega'|P\omega)
  =
  \zeta^P \gamma^{j'|j}_{q'|q}(\omega'|\omega),
\end{equation}
so does the sum of (b) and (d). The sum of (a) and (d) satisfies
\begin{equation}
  \label{eq:gamma_permute_spec3}
  \gamma^{Pj'|Pj}_{Pq'|Pq}(P\omega'|P\omega)
  =
  \gamma^{j'|j}_{q'|q}(\omega'|\omega),
\end{equation}
as does the sum of (b) and (c). Only the sum of all four diagrams
fulfils the complete relation~\eref{eq:gamma_permute}. In practice one
will exploit the symmetry imposed by~\eref{eq:gamma_permute} to reduce
the number of components which have to be evaluated directly. For
instance, it would be sufficient to calculate one of the four diagrams
in figure~\ref{fig:approx} and use the result to derive the other
three diagrams via~\eref{eq:gamma_permute}.

Since complex conjugation of a Green function interchanges creation
and annihilation operators it reverses the direction of lines in
diagrams. This results in the following criterion for a diagrammatic
approximation to comply with equation~\eref{eq:GF_conj_cont}
or~\eref{eq:vert_conj_cont} [or equivalently~\eref{eq:GF_conj}
or~\eref{eq:vertex_conj}]: the set of diagrams taken into account has
to be invariant under reversion of the directed lines combined with
the simultaneous exchange of the external indices $j$ with $j'$, $q$
with $q'$, and $\omega$ with $\omega'$. For example, reversing the
direction of all lines in figure~\ref{fig:approx}~(a) and
interchanging $1$ with $1'$ and $2$ with $2'$ reproduces the same
diagram. Hence, an approximation to the two-particle vertex function
consisting only of this single diagram complies
with~\eref{eq:vert_conj_cont}. The same holds for
figure~\ref{fig:approx}~(d). The diagram in
figure~\ref{fig:approx}~(b) is mapped onto that of
figure~\ref{fig:approx}~(c) by complex conjugation and vice versa. The
approximation given by their sum
fulfils~\eref{eq:vert_conj_cont}. Similarly to the rules for
permutation of particles, one will use~\eref{eq:GF_conj_cont}
or~\eref{eq:vert_conj_cont} in practice in order to reduce the number
of components one has to evaluate directly.

The theorem of causality for the vertex function has been proven in
section~\ref{subsec:causality} for each single diagram; the theorem of
analyticity is a direct consequence of it. Therefore, both theorems
hold in any diagrammatic approximation to the vertex functions. Given
the causal features of the noninteracting single-particle Green
function, the theorem of causality for the interacting multi-particle
Green function can also be proven to hold for any single diagram. The
proof is analogous to the one for the diagrams of vertex functions
given in section~\ref{subsec:causality}. That's why the theorems of
causality and analyticity for the Green functions hold in any
diagrammatic approximation as well.

The proof of the KMS condition~\eref{eq:gen_fluc_diss_vert} for the
vertex functions has been performed diagram per diagram. The same
applies to the special form~\eref{eq:gen_fluc_diss_spec_vert} which
brings about~\eref{eq:fluct_diss_gen_vert_Re}
and~\eref{eq:fluct_diss_gen_vert_Im}. Hence~\eref{eq:gen_fluc_diss_vert}
and~\eref{eq:fluct_diss_gen_vert_Re},~\eref{eq:fluct_diss_gen_vert_Im}
hold in any diagrammatic approximation to the vertex functions. Again,
an analogous proof for the diagrams of the interacting Green function
can be formulated, showing that equations~\eref{eq:gen_fluc_diss},
\eref{eq:fluct_diss_gen_Re}, and~\eref{eq:fluct_diss_gen_Im} hold in
any diagrammatic approximation to the Green functions.

\subsection{Functional renormalization group}

While diagrammatic approximations automatically preserve causality
and the KMS conditions, these features can be much more difficult to
maintain in other approximation schemes. We briefly sketch
complications for the fRG approach which is based on introducing a
flow parameter $\Lambda$ into the free single-particle Green
functions; typically this is done in such a way that degrees of
freedom at energies below $\Lambda$ are suppressed in order to
regularize low energy divergencies which may arise in perturbation
theory. The vertex functions, expressed as functionals of the free
single-particle Green functions, depend on $\Lambda$ and can be
computed as solutions of a coupled set of flow
equations \cite{wetterich, morris, salmhofer2}.

Let us consider the flow of the one-particle irreducible vertex
functions $\gamma_{n;\Lambda}$ governed by the hierarchy of the fRG
equations
\begin{eqnarray}
  \fl
  \frac{\rmd}{\rmd \Lambda} \gamma_{1;\Lambda}^{\nu'|\nu} = \zeta
  \gamma_{2;\Lambda}^{\nu' \lambda' | \nu \lambda} S_{\Lambda;
    \lambda|\lambda'} , \label{eq:frg1} 
  \\ 
  \fl
  \frac{\rmd}{\rmd \Lambda} \gamma_{2;\Lambda}^{\nu'_1 \nu'_2 | \nu_1 \nu_2} 
  = \zeta \gamma_{3;\Lambda}^{\nu'_1 \nu'_2 \lambda' | \nu_1 \nu_2 \lambda} 
  S_{\Lambda; \lambda|\lambda'} 
  +  \gamma_{2; \Lambda}^{\nu'_1 \nu'_2 | \lambda_1 \lambda_2}  G_{\Lambda;
    \lambda_1| \lambda'_1} S_{\Lambda; \lambda_2 | \lambda'_2}
  \gamma_{2;\Lambda}^{\lambda'_1 \lambda'_2 | \nu_1 \nu_2}
\nonumber \\
\fl
\qquad
+ \left\{ G_{\Lambda; \lambda_1 | \lambda'_2} S_{\Lambda; \lambda_2 |
  \lambda'_1} \left[ \zeta \gamma_{2;\Lambda}^{\nu'_1 \lambda'_1 |
    \nu_1 \lambda_1}  \gamma_{2; \Lambda}^{\lambda'_2 \nu'_2 |
    \lambda_2 \nu_2} + \gamma_{2;\Lambda}^{\nu'_1 \lambda'_1 |
    \lambda_1 \nu_2}  \gamma_{2; \Lambda}^{\lambda'_2 \nu'_2 | \nu_1
    \lambda_2} \right] 
+ (G_{\Lambda} \leftrightarrow S_{\Lambda}) \right\},
\label{eq:frg2} \\
\fl
\ldots,
\label{eq:frg3}
\end{eqnarray}
where $\gamma_{1;\Lambda} \equiv \Sigma_{\Lambda}$, and the
multi-indices $\nu$ and $\lambda$ comprise frequency (time) argument
as well as the state and the Keldysh indices; a summation convention
applies to multi-indices appearing twice in a product. The dots
in~\eref{eq:frg3} represent the flow equations for
$\gamma_{3;\Lambda}, \gamma_{4;\Lambda}, \ldots$. The notation $S_{\Lambda}$ stands for
the so-called single scale propagator which is defined by
\begin{equation}
S_{\Lambda; \lambda | \lambda'} = - G_{\Lambda; \lambda | \nu} \left(
  \frac{\rmd (G_{\Lambda}^{(0)})^{-1}}{\rmd \Lambda} \right)^{\nu | \nu'}
G_{\Lambda; \nu' | \lambda'} . 
\label{eq:singlescale}
\end{equation}
The equations \eref{eq:frg1}~--~\eref{eq:frg3} can be systematically
obtained from the vertex generating functional explicitly depending of
the flow parameter $\Lambda$ \cite{salmhofer2, gezzi}. An
equivalent derivation is based on a purely diagrammatic approach
\cite{jakobs1} in which $\Lambda$ is explicitly introduced into the
free propagators $G^{(0)}$.

Since the flow of $\gamma_{n;\Lambda}$ is determined by
$\gamma_{1;\Lambda}, \ldots, \gamma_{n;\Lambda}$ and $\gamma_{n+1;
  \Lambda}$, the exact flow equations~(\ref{eq:frg1}--\ref{eq:frg3})
form an infinite coupled hierarchy. The final solution of the full
hierarchy is exact and therefore satisfies all relations discussed in
the previous sections. However, in nontrivial practical problems one
cannot solve the full hierarchy but has to resort to approximations.
Typically one truncates the hierarchy of flow equations which means
setting $\rmd \gamma_{n;\Lambda} / \rmd \Lambda = 0$ for some $n$,
such that the flow equations for $\gamma_{1;\Lambda}, \ldots,
\gamma_{n-1;\Lambda}$ form a finite closed set. Then it is no longer
guaranteed that the final solution of the truncated flow equations
satisfies the exact relations in question. It turns out that the
choice of the flow parameter has a decisive influence on their
conservation. While arbitrary choices of the flow parameter do not
affect the properties under permutation of particles and complex
conjugation, special care is required in order to maintain the theorem
of causality and the KMS conditions.

Assuming that the only approximation done to the flow equation is a
truncation of the hierarchy we find the following sufficient
conditions for the construction of conserving flow parameters. (1) If
the free propagator $G^{(0)}_\Lambda$ furnished with the flow
parameter $\Lambda$ satisfies the theorem of causality, then the
vertex functions resulting from a truncated set of flow equations
fulfil the theorem of causality throughout the flow and in the final
solution. (2) If $G^{(0)}_\Lambda$ satisfies the fluctuation
dissipation theorem~\eref{eq:fluct_diss} (with the factor $\left[1 +
  \zeta 2 n_\zeta (\omega-\mu) \right]$ not depending on $\Lambda$)
then the vertex functions resulting from a truncated set of flow
equations fulfil the KMS conditions throughout the flow and in the
final solution.

For the proof of these rules we recall that the theorem of causality
and the KMS conditions for the vertex functions have been derived in
sections~\ref{subsec:causality} and~\ref{subsec:gen_fluc_diss} for
each single contributing diagram on the basis of the corresponding
property of the free propagator $G^{(0)}$. In reference~\cite{jakobs1}
the flow equations for the vertex functions have been derived from the
individual diagrams contributing to $\rmd \gamma_{n;\Lambda}/ \rmd
\Lambda$, which contain lines $G^{(0)}_\Lambda$ and one line $\rmd
G^{(0)}_\Lambda / \rmd \Lambda$. If $G^{(0)}_\Lambda$ and $\rmd
G^{(0)}_\Lambda / \rmd \Lambda$ satisfy the theorem of causality, then
the same arguments as applied in section~\ref{subsec:causality} can be
used to prove the theorem of causality for each single diagram
contributing to $\rmd \gamma_{n;\Lambda}/ \rmd \Lambda$. Truncating
flow equations means taking into account only a subset of these
diagrams and it thus has no influence on the conservation of causality
during the flow. Note that the theorem of causality for $\rmd
G^{(0)}_\Lambda / \rmd \Lambda$ follows from that of $G^{(0)}_\Lambda$
since the derivative is a linear operation. Concerning the KMS
conditions we can argue analogously.

As examples for these considerations we discuss different flow
parameters used in the literature and propose also a new one. In
reference~\cite{gezzi} the flow parameter is introduced into the free
Green function by setting
\begin{equation}
  G^{(0)}_\Lambda(\omega)
  =
  \Theta(|\omega|-\Lambda) G^{(0)}(\omega).
\end{equation}
Since the step function $\Theta(z)$ has a branch-cut along the
imaginary axis of $z$, the analytic properties of
$G^{(0)}_\Lambda(\omega)$ are in contradiction to the theorem of
causality: $G^{(0)}_\Lambda(\omega)$ being nonanalytic in the upper
half plane implies that $G^{(0) \, \ix{Ret}}_\Lambda(t|t') = G^{(0) \,
  2|1}_\Lambda(t|t')$ does not vanish for $t' > t$. As a consequence
the vertex functions constructed from diagrams with lines given by
$G^{(0)}_\Lambda$ instead of $G^{(0)}$ do not exhibit the correct
causal features. When the exact RG flow stops at $\Lambda = 0$,
causality is restored due to $G^{(0)}_{\Lambda = 0} =
G^{(0)}$. However, for nontrivial models the flow equations for the
vertex functions can be solved only approximately. Restoring causality
for approximate solutions of the flow equations was encountered as a
major difficulty in reference~\cite{gezzi}.

This problem can be solved if one chooses a flow parameter such that
$G^{(0)}_\Lambda(\omega)$ has the same analytic features as
$G^{(0)}(\omega)$. Then the causal features of the vertex functions
are preserved during all the RG flow, even for truncated flow
equations. An example of such a flow parameter is the imaginary
frequency cut-off proposed in reference~\cite{jakobs1} which
manipulates the particle distribution functions of the reservoirs by
\begin{equation}
  \label{eq:imag_freq_cut-off}
  \fl
  n_\zeta(\omega)
  =
  \frac{1}{\rme^{\beta \omega} - \zeta}
  =
  T \sum_{\omega_n} \frac{\rme^{\rmi \omega_n 0^+}}{\rmi
    \omega_n - \omega}
  \qquad  \rightarrow \qquad
  n_\zeta^\Lambda(\omega)
  =
  T \sum_{\omega_n} \frac{\Theta(|\omega_n| -\Lambda) \rme^{\rmi
      \omega_n 0^+}}{\rmi \omega_n - \omega},
\end{equation}
where $\omega_n = 2 n\pi T$ or $\omega_n = (2n+1)\pi T$ denote the
Matsubara frequencies for bosons or fermions, respectively. This
cut-off affects only the Keldysh component of the free single-particle
Green function, but not the retarded or advanced ones which contain
only dynamical but no statistical information, $G^{(0)
  \ix{Ret/Av}}_\Lambda(\omega) = G^{(0) \ix{Ret/Av}}(\omega)$. Hence,
the causal properties of free single-particle Green function are
unchanged and causal features of the vertex functions are conserved.
However, since the $\Lambda$-dependent distribution function
in~\eref{eq:imag_freq_cut-off} is not the thermal one, the
single-particle Green function $G^{(0)}_\Lambda$ does not fulfil the
fluctuation dissipation theorem for $\Lambda \neq 0$. As a consequence
the KMS conditions for the vertex functions are not satisfied until
the exact flow terminates at $\Lambda=0$. Solvable approximations to
the flow equations will in general not restore the KMS-conditions even
at $\Lambda=0$. (The simple static approximation used in
reference~\cite{jakobs1} yet satisfies the KMS-conditions in a trivial
way.)

Examples of flow parameters which violate neither causality nor the
KMS conditions are given by the momentum cut-off and by the
hybridization of the system and reservoirs. 

The momentum cut-off is introduced for extended systems by defining
\begin{equation}
G^{(0) \ix{Ret/Av/K}}_{k;\Lambda} (\omega) = \Theta (|\varepsilon_k| -\Lambda) 
G^{(0) \ix{Ret/Av/K}}_k (\omega), 
\label{eq:gblam}
\end{equation}
where the wave number $k$ labels a single-particle eigenstate with
eigenenergy $\varepsilon_k$. Obviously, this cut-off does not affect
the frequency dependence of $G^{(0)}$; hence, it preserves causality
and the KMS conditions.

We propose to use hybridization as a flow parameter which is useful
for local systems coupled to external reservoirs.  After integrating
out reservoir states, the Green function of a local system acquires a
reservoir broadening $\Gamma$. For example, the reservoir-dressed
Green function of the SIAM defined in \eref{eq:H_B} reads
\begin{equation}
  G^{(0) \ix{Ret}}_\sigma (\omega) = \frac{1}{\omega - (\varepsilon_0 +\sigma
    B) + \frac{\rmi}{2} \Gamma}, 
\label{eq:ganderdef}
\end{equation}
where $\Gamma = 2 \pi \sum_{p} \delta (\mu - \varepsilon_p) |V_p
|^2$. The idea of the hybridization flow parameter consists in
introducing an additional -- fictitious -- reservoir, which is
assumed to be in thermodynamical equilibrium with the physical one,
i.e. is characterized by the same temperature $T$ and chemical
potential $\mu$. The flow parameter $\Lambda$ is then associated
with an additional broadening due to a coupling to the fictitious
reservoir, so that
\begin{eqnarray}
  G_{\sigma;\Lambda}^{(0) \ix{Ret}} (\omega) &= \frac{1}{\omega -
    (\varepsilon_0 +\sigma B) + \frac{\rmi}{2} (\Gamma +\Lambda)},
  \label{eq:g0lamRet}
  \\
  G_{\sigma;\Lambda}^{(0) \ix{Av}} (\omega) &= G_{\sigma;\Lambda}^{(0)
    \ix{Ret}} (\omega)^\ast,
  \label{eq:g0lamAv}
  \\
  G_{\sigma;\Lambda}^{(0) \ix{K}} (\omega) &=
  \left[1 + \zeta 2 n_\zeta (\omega-\mu) \right]
  \left[G_{\sigma;\Lambda}^{(0) \ix{Ret}}(\omega)
    -G_{\sigma;\Lambda}^{(0) \ix{Av}}(\omega) \right].
  \label{eq:g0lamK}
\end{eqnarray}
In the beginning of the flow at $\Lambda = \infty$, the free
propagator is suppressed and the vertex functions are identical to
their bare values (except for the self-energy which acquires a final
contribution from the Hartree-Fock diagram, $\Sigma^{\ix{Ret}}_{q'|q;
  \Lambda=\infty} = \frac{1}{2} \sum_p \overline v_{q'p|qp}$). In the
end of the flow at $\Lambda=0$ the fictitious reservoir is decoupled,
and we restore the original system. Since $G^{(0)}_\Lambda$ in
(\ref{eq:g0lamRet}--\ref{eq:g0lamK}) satisfies the theorem of
causality and the fluctuation dissipation theorem, causality and the
KMS conditions of the vertex functions are preserved during the flow,
even when the flow equations are truncated. A detailed study of the
SIAM on basis of the hybridization flow is presented
in~\cite{jakobs2}, where it is also generalized to nonequilibrium
situations.

%====================================================================

%====================================================================

\section{Conclusion}

Currently the interest in nonequilibrium dynamics and in the crossover
between equilibrium and nonequilibrium properties is enormously
increasing in the field of condensed matter physics. Keldysh formalism
provides the possibility to treat equilibrium and nonequilibrium
quantum systems on equal footing. Numerous equilibrium techniques
based on Green or vertex functions benefit a lot from adaptation to
Keldysh formalism. For its correct implementation the knowledge of
real time (frequency) features of the multi-component Green and vertex
functions appears to be mandatory. The present review lists general
properties of these functions in equilibrium and nonequilibrium
stationary state.

The relations for the Green functions have been derived directly from
their definition as expectation value of contour ordered operator
product. The corresponding equations for the vertex functions have
been obtained from the rules for the evaluation of irreducible
diagrams.

We have surveyed the relations for permutations of
particles~(\ref{eq:G_permute},\,\ref{eq:gamma_permute}) and complex
conjugation~(\ref{eq:GF_conj_cont},\,\ref{eq:vert_conj_cont}) and cast
them into the form convenient in practical use. In
section~\ref{subsec:causality} we have formulated the causality
theorem and identified the components of the Green and vertex
functions which are analytic in a certain half plane of the
corresponding frequency arguments. These analytic features should be
maintained throughout any approximation to the Green and vertex
functions, otherwise the fundamental property of causality is
violated.

A major part of the review has been devoted to the study of the KMS
conditions which hold in equilibrium. We have worked out the
constraints which equilibrium imposes on the multi-particle functions.
They are given by the
relationships~(\ref{eq:gen_fluc_diss},\,\ref{eq:gen_fluc_diss_vert})
to the multi-particle functions of the time reversed system. The
connection to the FDT for single-particle functions has been
extensively discussed. In particular we have determined the reason for
its special form~\eref{eq:fluct_diss} for $n=1$: it follows from the
peculiar single-particle property~\eref{eq:G_G_tilde}. Furthermore, we
identified an important class of systems with special
behaviour~\eref{eq:time_rev_inv_GF} under time reversal which allow
for a simplified
formulation~(\ref{eq:fluct_diss_gen_Re},\,\ref{eq:fluct_diss_gen_Im},\,\ref{eq:fluct_diss_gen_vert_Re},\,\ref{eq:fluct_diss_gen_vert_Im})
of the multi-particle FDT.

Finally we discussed how the properties in question relate to
diagrammatic approximations and to approximations within the fRG. For
diagrammatic approximations we found that the set of diagrams taken
into account should be invariant under permutations within the
incoming and within the outgoing external indices and invariant under
reversal of the directed lines. Then the relations for permutation of
particles and complex conjugation hold and reduce considerably the
number of independent components which simplifies a treatment of a
problem. Causal features and KMS conditions are preserved
automatically in any diagrammatic approximation. However,
approximations to the flow equations of the fRG often destroy
causality and the KMS conditions. These properties can be preserved
when the flow parameter is appropriately chosen in such a way that it
respects the corresponding properties of the noninteracting
single-particle Green function.

%====================================================================

\ack

SGJ. thanks F~Reininghaus, U~Heinz and E~Wang for helpful
discussions. This work was supported by the DFG-Forschergruppe~723.

%====================================================================

%====================================================================

%====================================================================

%====================================================================

\end{document}